\documentclass[sigconf]{acmart}
\AtBeginDocument{%
  }

\copyrightyear{2025}
\acmYear{2025}
\setcopyright{acmlicensed}\acmConference[UIST '25]{The 38th Annual ACM Symposium on User Interface Software and Technology}{September 28-October 1, 2025}{Busan, Republic of Korea}
\acmBooktitle{The 38th Annual ACM Symposium on User Interface Software and Technology (UIST '25), September 28-October 1, 2025, Busan, Republic of Korea}
\acmDOI{10.1145/3746059.3747722}
\acmISBN{979-8-4007-2037-6/2025/09}

\usepackage{amsmath}
\usepackage{xcolor}
\usepackage{soul} 
\usepackage{booktabs}

\makeatletter
\patchcmd{\@printtopmatter}
 {0.9\textheight}
 {1.0\textheight}
 {}{}
\makeatother

\newcommand{\system}{\textsc{Gumbo}}
\newcommand{\model}{\textsc{GUM}}
\newcommand{\rev}[1]{{#1}}

\begin{document}
\title{Creating General User Models from Computer Use}

\author{Omar Shaikh}
\affiliation{%
  \institution{Stanford University}
  \city{Stanford}
  \country{USA}
}

\author{Shardul Sapkota}
\affiliation{%
  \institution{Stanford University}
  \city{Stanford}
  \country{USA}
}

\author{Shan Rizvi}
\affiliation{%
  \institution{Independent}
  \city{New York City}
  \country{USA}
}

\author{Eric Horvitz}
\affiliation{%
  \institution{Microsoft Research}
  \city{Redmond}
  \country{USA}
}

\author{Joon Sung Park}
\affiliation{%
  \institution{Stanford University}
  \city{Stanford}
  \country{USA}
}

\author{Diyi Yang}
\affiliation{%
  \institution{Stanford University}
  \city{Stanford}
  \country{USA}
}

\author{Michael S. Bernstein}
\affiliation{%
  \institution{Stanford University}
  \city{Stanford}
  \country{USA}
  }

\renewcommand{\shortauthors}{Shaikh et al.}

\begin{abstract}
Human-computer interaction has long imagined technology that understands us---from our preferences and habits, to the timing and purpose of our everyday actions. Yet current user models remain fragmented, narrowly tailored to specific applications, and incapable of the flexible, cross-context reasoning required to fulfill these visions. This paper presents an architecture for a \textit{general} user model (\model{}) that learns about you by observing any interaction you have with your computer. The \model{} takes as input any unstructured observation of a user (e.g., device screenshots) and constructs confidence-weighted natural language propositions that capture that user's behavior, knowledge, beliefs, and preferences. \model{}s can infer that a user is \texttt{preparing for a wedding they’re attending} from a message thread with a friend. Or recognize that a user is \texttt{struggling with a collaborator’s feedback} on a draft paper by observing multiple stalled edits and a switch to reading related work. \model{}s introduce an architecture that infers new propositions about a user from multimodal observations, retrieves related propositions for context, and continuously revises existing propositions. To illustrate the breadth of applications that \model{}s enable, we demonstrate how they augment chat-based assistants with contextual understanding, manage OS notifications to surface important information only when needed, and enable interactive agents that adapt to user preferences across applications. We also instantiate a new class of proactive assistants (\system{}s) that discover and execute useful suggestions on a user's behalf based on their GUM. In our evaluations, we find that \model{}s make calibrated and accurate inferences about users, and that assistants built on \model{}s proactively identify and perform actions of meaningful value that users wouldn’t think to request explicitly. Altogether, \model{}s introduce new methods that leverage large multimodal models to understand unstructured user context, enabling both long-standing visions of HCI and entirely new interactive systems that  anticipate user needs.
\end{abstract}


\begin{teaserfigure}
  \centering
  \includegraphics[width=0.96\textwidth]{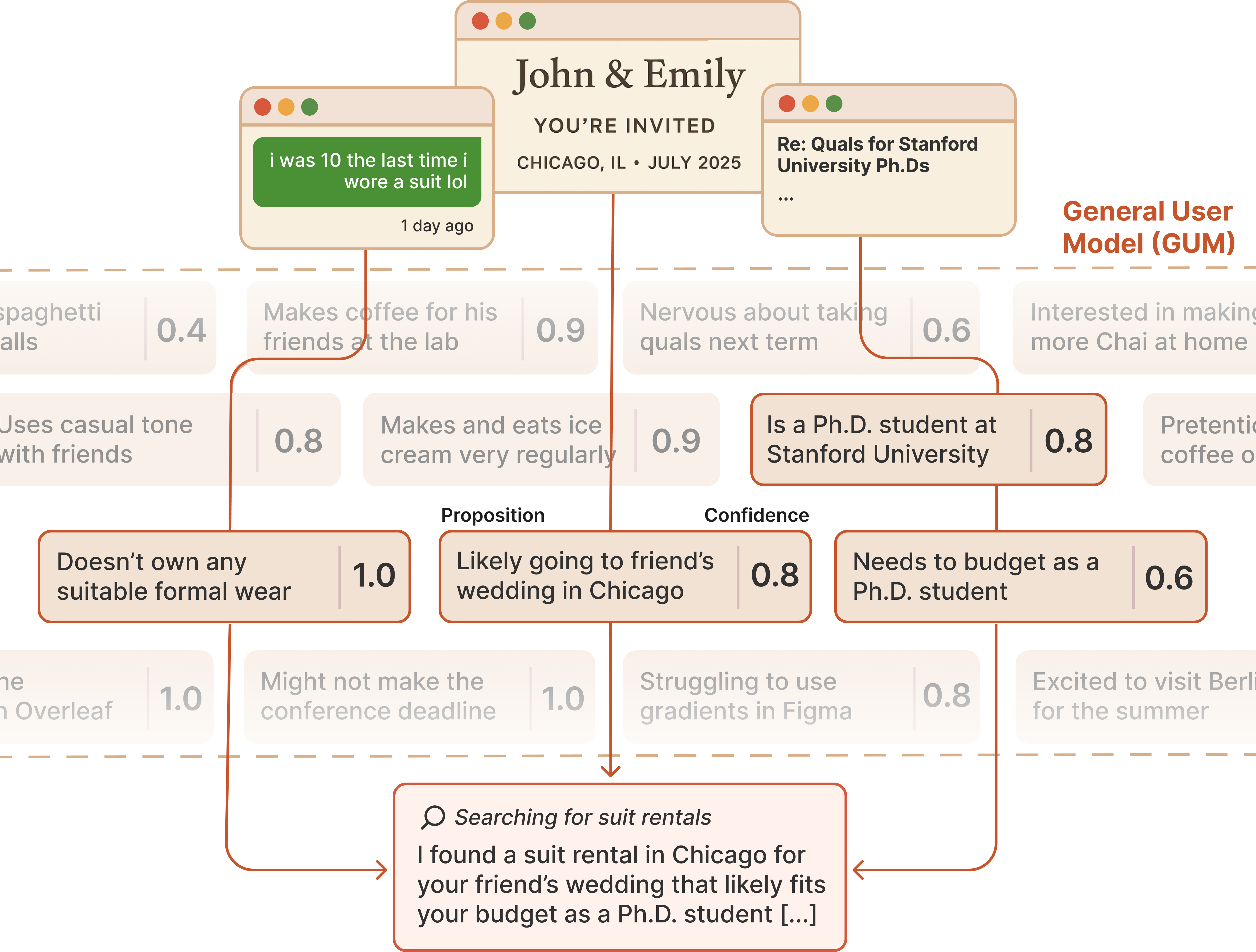}
  \caption{General User Models take as input completely unstructured interaction data (top), transforming these observations into structured propositions about a user (center). Together, these propositions create a \emph{general user model} (\model{}). We instantiate our user model in an assistant (\system{}) that proactively discovers and executes suggestions on a user's behalf (bottom).}
  \label{fig:teaser}
\end{teaserfigure}


\maketitle
\section{Introduction}

In our most celebrated visions of human-computer interaction, interactive systems deploy a rich understanding of our goals, tasks, and context. Mark Weiser's central scenario in \emph{The Computer For the 21st Century} opens with an alarm clock that knows when its owner is about to wake, and proactively offers coffee~\cite{weiser1991computer}. Likewise, Apple's Knowledge Navigator looks up useful supporting information while its user is puzzling over a question, then blocks an undesired phone call while the user is focused~\cite{apple1987knowledge}. This notion of technology that \textit{knows the user well enough to do the right thing at the right time} weaves through visions of both context-aware systems~\cite{dey2001understanding}, which promise to adapt as the user's situation evolves, and interface agents~\cite{maes1993learning}, which take proactive action on the user's behalf.

Today, however, these visions remain largely out of reach. Despite progress in user modeling, recommender systems, and context-aware systems, computers remain remarkably unaware of who we are, what we are doing, and what would be helpful. At their core, current user models are simply too \textit{narrow}: they understand our music preferences, our use of tools within a single app, or the TV show we might watch next. Even when user models integrate data across multiple applications, the integration remains surface-level; user models cannot reason or make inferences in new contexts. Our visions of technology, however, require user models that are \textit{broad}, able to reason about everything from our general preferences to our immediate information needs; and capable of applying these insights across contexts, from work-related tasks to recreational activities. Applications today stumble because they have pinhole views of users: Weiser's vision of ubiquitous computing requires models that reason about family, friends, and work---not one application, and not just via a one-dimensional signal.

In this paper, we describe an architecture for \textit{\textbf{general user models}}: computational models that materialize information and inferences about a user across domains and time scales.\footnote{A demo of \system{} and our open source package for \model{} are available at \url{https://generalusermodels.github.io}} The general user model, or \textbf{GUM}, allows a user to construct a private, computational representation of their own behavior,
knowledge, beliefs, and preferences by feeding unstructured observations (e.g., screenshares of their computer use) through an inference architecture leveraging large multimodal models (e.g., vision and language models, or VLMs). Our architecture contributes three main elements. A \textit{Propose} module translates unstructured observations into confidence-weighted propositions about the user's preferences, context, and intent. A \textit{Retrieve} module indexes and searches these propositions to return the most contextually relevant subset for a given query. Finally, using results from Retrieve, a \textit{Revise} module reevaluates and refines propositions as new observations arrive. We audit all observations for privacy violations with a contextual integrity~\cite{nissenbaum2004privacy} \emph{Audit} module that leverages the GUM itself to estimate and filter out information that the user would not expect to be recorded into the GUM. All data stays securely on the user's device, enabling local inference on capable hardware.

The operating system, applications, or the user themselves can query the GUM in real time to realize a breadth of applications similar to those envisioned in foundational HCI work. As part of the \model{}, we introduce an interface that enables applications to query the GUM for underlying propositions. Any unstructured observation that the \model{} sees can be marshalled to power interactive applications. Regardless of the interaction, users maintain direct and local control of the \model{}'s underlying propositions, allowing for edits, deletions, or additions.

At the simplest level, the GUM can insert information to establish common ground between applications and a user: for example, automatically adding relevant context when prompting a language model such as ChatGPT. With a \model{}, any LLM can now directly reference the research paper you were reading just minutes earlier when you ask about its methodology, eliminating the need for you to explicitly quote or summarize the paper's content. Beyond prompting LLMs, any application can directly query the GUM to adapt their experiences, realizing long-standing HCI visions. A GUM-enhanced operating system, for example, could prioritize only truly relevant notifications during a meeting---surfacing an imminent conference registration deadline while suppressing recipe emails. Email clients connected to a GUM could automatically sort messages based on observed user priorities, without requiring additional application-specific training. 

\model{}s also enable the creation of an entirely new class of proactive, interactive systems. We demonstrate this through an assistant, \textbf{\system{}}, that learns a \model{} via continuous private screenshot capture of the user's computer. Using the \model{}, \system{} constantly discovers \emph{what} suggestions would be helpful to surface conditioned on the user's context. In addition, \system{} uses the underlying \model{} to determine \emph{if and when} it might be useful to intervene with, and autonomosuly execute on, a suggestion. By marshalling a user's context, \system{} can proactively discover a range of useful suggestions and filter important ones appropriately. 

For the first author of this paper, \system{} proactively found a location to rent a suit after observing a wedding invite from their friend (constrained by the author's budget, see Fig \ref{fig:teaser}). \system{} also found and proposed fixes for bugs in the system \emph{itself} during development; and suggested potential revisions to this paper based on interactions with collaborators. For participants in our evaluation, \system{} brainstormed ways to integrate a new theoretical framework into ongoing research, created a highly personalized moving plan for a cross-country relocation, and helped organize email archives from scattered correspondence---all proactively, based solely on its observations of users.  

In our technical evaluations, we first focus on validating \model{} accuracy. We train \model{} on recent email interaction, feeding each email---metadata, attachments, links, and replies---sequentially into the \model{}. \rev{Emails contain diverse and relevant information across modalities, while also being full of irrelevant content. This makes synthesizing important information difficult: \model{}s must maintain accuracy despite this challenge. } Still, $N=18$ participants judged propositions generated by \model{}s as overall accurate and well-calibrated: unconfident when incorrect, and confident when correct. Highly confident propositions (confidence = 10) were rated 100\% accurate, while all propositions on average---including ones with low confidence---were fairly accurate (76.15\%). From ablation studies, we show that all \model{} components are critical for accuracy. We then deploy \system{} with $N=5$ participants for 5 days, with the system observing the participants' screens. This longitudinal evaluation replicated our results with the underlying \model{}. Additionally, participants identified a meaningful number of useful and well-executed suggestions completed by \system{}. Two of the five participants found particularly high value in the system and asked to continue running it on their computer after the study concluded. Our evaluations also highlight limitations and boundary conditions of \model{} and \system{}, including privacy considerations and overly candid propositions.

In sum, we contribute \textit{General User Models (\model{}s)}: computational representations of a user's behavior, knowledge, beliefs, and preferences, built from unstructured observations of the user. We demonstrate an implementation of a \model{}, an interface allowing applications to query the \model{}, an example assistant application called \system{}, a technical evaluation via unstructured email interactions, a longitudinal evaluation via unstructured screen captures, and a reflection on norms and implications of this class of application.
\section{Related Work}

The first major challenge that \model{}s must engage with is common ground: the mutual knowledge, beliefs, and assumptions shared between individuals that enables efficient communication \cite{clark1989contributing}. Through dialogue, gestures, facial expressions, and other multimodal cues \cite{clark1996using}, people continually update shared mental models, iteratively constructing common ground \cite{clark1991grounding}. Common ground acts as a shared context, allowing people to communicate without unnecessary elaboration.

However, AI systems have little shared context with users and make assumptions about a user’s background, producing one-size-fits-all answers that are over-informative \cite{tsvilodub2023overinformative}, overconfident \cite{mielke2022reducing}, and unable to handle ambiguity \cite{abercrombie2023mirages, min2020ambigqa, gao2021answering}. Inability to ground extends beyond question-asking. For example, designing and prototyping AI systems requires collaborative grounding \cite{vaithilingam2024imagining}. Otherwise, they might take unfavorable and irreversible action~\cite{bansal2024challenges, shao2024collaborative, cemri2025multi}.

To bridge the human-AI grounding gap, two general solutions have emerged: asking clarification questions or learning from a user’s dialogue history. Clarification-based grounding attempts to simulate the natural back-and-forth dialogue that humans use to establish shared understanding. This interaction can be implemented through prompting \cite{kuhn2022clam, chen-etal-2023-controllable}, finetuning \cite{andukuri2024star, zhang2023clarify, hong2023zero, gan2024clarq}, or a combination of both \cite{shaikh2025navigating}. 
Keeping a \rev{long-term} memory of a user’s chat interaction history is another solution \cite{lucas2009managing, joshi2017personalization, zhang2018personalizing, packer2023memgpt, shaikhaligning}. \rev{Long-context models also allow users to provide lengthy inputs as grounding~\cite{team2024gemini}, but models generally forget information as context lengths increase~\cite{liu2023lost}. Meanwhile, retrieval-augmented generation (RAG)~\cite{lewis2020retrieval}---where context is retrieved from a traditional database---has served as an alternative approach. For example, systems like OmniQuery~\cite{li2025omniquery} draw from a processed database of multimodal memories from photo albums, contextualizing queries to an LLM.}  Regardless of the approach, current models are limited to context derived solely from their own prior interactions with the user \rev{or from a limited taxonomy of pre-defined applications or sources.} All other computer interaction is inaccessible to models.

We argue that engaging in grounding only when a user interacts with a model is a constraining interaction paradigm. Piecing together custom grounding interactions for specific domains---and fixing these interactions to dialogue---limits the ubiquity of potential common ground users can build with AI systems. In this work, we propose \emph{general} user models that proactively build common ground with users through observation.

The second major issue that \model{}s draw on is that prompting LLMs effectively is difficult~\cite{zamfirescu2023johnny}. Users often provide models with underspecified prompts~\cite{zamfirescu2023herding, shaikh2025navigating, agrawala2023unpredictable}, requiring repeated iteration to specify constraints for a specific context. A range of interactive systems have enabled users to better specify constraints to LLMs: systems like ChainForge~\cite{arawjo2024chainforge} and EvalGen~\cite{shankar2024validates} offer users a means to interactively validate and iterate on prompts for specific tasks. Beyond developing prompts, systems like Graphologue~\cite{jiang2023graphologue} and Sensescape~\cite{suh2023sensecape} allow users to explore the range of possible outputs given a prompt. Across these interactions, the burden of providing explicit input to build shared context lies with the user. 

In contrast, a longstanding goal in HCI is to build systems that proactively assist users without disruption. Early work in mixed-initiative interfaces balanced autonomy with user control, leveraging context to identify the best moments to interact~\cite{horvitz1999principles}. Classic AI models have already been employed to understand a user's multimodal context. Memory Landmarks and the LifeBrowser system~\cite{horvitz2004learning}, for example, construct an understanding of which events users find important~\cite{ringel2003milestones} by analyzing their calendar and mailbox. A handful of systems have started using LLMs to selectively process context. PICAN~\cite{hong2024context}, for example, focuses on building context in VR-onboarding settings by focusing on dialogue and actions the user takes in the metaverse. Systems like GPTCoach~\cite{jorke2024supporting} rely on external health data and qualitative context, augmenting LLMs with selective context to help with behavior change. 

Still, our interactions with LLMs are far from visions like Weiser's Sal, where computers unify machine intelligence with context from our daily lives~\cite{horvitz2006machine}. Instead, LLM-based systems focus on \emph{constraining} the types of inputs, outputs, and tasks models operate over. In this paper, we embrace what LLMs are good at---unconstrained input---and ground assumptions about the user and their context based on observation. We design what is effectively a multimodal grounding engine. 

Context aware computing has long envisioned such toolkits~\cite{schilit1994context, dey2001understanding, dey2001conceptual}. Yet these frameworks are typically pinned down to some underlying fixed \emph{structure} of the input data or the underlying sensors~\cite{dourish2004we}. This inevitably limits how we define context and the types of applications we can support from the outset. \model{}s offer an alternative approach---they introduce a flexible structure with very few constraints to build context.

\section{General User Models}
We introduce an architecture for learning general user models. \model{}s enable learning both stable inferences about the end user and moment-to-moment context, using everyday action as input data. From low-level input (e.g. screenshots), \model{}s construct a rich understanding of who we are and what we are doing. Applications can then query the \model{} to power a range of interactions, from operating systems that power contextual notifications to assistants that proactively discover and execute useful suggestions based on context.

At its core, \model{} is a collection of natural language, confidence-weighted propositions about a user, learned entirely from observation. The following propositions, for example, were constructed by the author's \model{} while they wrote this section. Each proposition is also paired with a confidence score (from 0 to 1).
\begin{quote}
    \textbf{Proposition}: \texttt{Omar is currently writing in the General User Model's section.}\\
    \textbf{Confidence}: 0.8\\
    \\
    \textbf{Proposition}: \texttt{Omar is viewing and resolving comments from their advisors, Diyi and Michael.}\\
    \textbf{Confidence}: 0.7\\
    \\
    \textbf{Proposition}: \texttt{Omar is struggling with the technical evaluation of \model{}s.}\\
    \textbf{Confidence}: 0.5
\end{quote}
Propositions were generated entirely using a series of screenshots from the author's screen. \model{} takes any type of input that an underlying large model can accept (in the case of vision-langauge models, \model{} accepts image and text input). Using this unstructured input, \model{} composes and refines proposition-confidence pairs. This abstraction---a series of confidence-weighed statements that conjecture some \textit{common ground} between the user and their computer---enables a flexible representation of a user's context. To be clear: \model{}s do not enforce any structure on the types or kinds of propositions that can be generated. Propositions are constrained only by natural language descriptions, and are revised as the \model{} continues receiving more input. \model{}s aim to continually learn this representation of a user's context from any interaction. 

Here, we detail abstractions (confidence-weighted propositions) that power the \model{} (\S\ref{sec:abstractions_detailed}) and document an interface for using \model{}s in other applications (\S\ref{sec:python_interface}).

\subsection{Propositions and Confidences: Abstractions for Representing General User Models}
\label{sec:abstractions_detailed}
\model{} draws inspiration from literature on common ground in human-human interaction~\cite{clark1991grounding, clark1996using}. A subset of common ground between participants can be defined by shared assumptions. The human-AI grounding challenge lies in identifying \emph{valid} assumptions an AI model can make about a user. When constructing a user model, we must include shared assumptions that accurately represent user actions. For example, when a user repeatedly searches for "best hiking trails in Colorado" and visits outdoor equipment websites, the model might form the proposition "User enjoys outdoor activities, and is interested in hiking in Colorado" with high confidence. 

We build abstractions for \model{} based on grounding models in human-human communication~\cite{brennan2014grounding}. Bayesian approaches represent shared knowledge as probabilistic distributions, not deterministic sets of propositions. In these models, interlocutors maintain probability distributions over possible states, refining a shared understanding through communication~\cite{paek1999uncertainty, paek2013conversation, goodman2016pragmatic}. This probabilistic view accommodates user uncertainty and grounding over time~\cite{lassiter2012presuppositions,horvitz2013lumiere}. To instantiate a probabilistic model of common ground, we leverage the in-context learning capabilities of large langauge models, which can be loosely interpreted as Bayesian updates~\cite{xie2021explanation, zhang2024should}.
We operationalize these approaches through three abstractions: observations, propositions, and confidences.

\textbf{Observations} are raw inputs that the \model{} processes. Observations can be anything, so long as they can be tokenized and fed into a large (vision/language/multimodal) model. Screenshot observations provide visual data of what the user is viewing, file system observations reveal document interactions, and notification observations offer communication pattern insights. Observations are the source from which \model{} builds user context understanding. Each observation includes metadata: the source, timestamp, and a unique identifier. Unlike propositions, observations are factual records, not inferences. 

\textbf{Propositions} are the assumptions, in text, that \model{} constructs through observation. Viewing a friend's wedding invite yields surface inferences like \texttt{User is invited to friend's wedding}. However, \model{} may generate more complex inferences requiring speculation, e.g., \texttt{User needs to buy or rent a suit for the wedding}. These propositions might not be correct, making uncertainty inclusion critical. We, therefore, include confidences alongside each proposition.

\textbf{Confidences} reflect how certain the model is about proposition truth, predicated on the quality and quantity of the available evidence to support the proposition. Confidences enable decision making over uncertainty, representing belief degrees associated with each proposition. This allows prioritizing actions based on reliable information while acknowledging tentative assumptions. \texttt{User is invited to friend's wedding} requires less speculation: the invite appears on-screen, earning high confidence (1.0). Inferences about \texttt{needing to buy or rent a suit} are never explicitly expressed, receiving lower confidence (0.4). In \S\ref{sec:tech_eval}, we evaluate how well the \model{}'s generated confidence scores align with participants' actual accuracy.

Additionally, propositions include \textbf{metadata} for downstream applications. Inspired by cognitive agent architectures~\cite{anderson1993rules, park2023generative}, we introduce generated \textbf{decay} scores (0-1), indicating proposition staleness rates. \texttt{Author is a Ph.D. student} decays slowly (score = 1.0). Conversely, \texttt{Author is debugging a memory leak} is transient (score = 0.6). Each proposition includes \textbf{grounding}: observations supporting the inference, plus generated \textbf{reasoning} explaining the proposition's construction:
\begin{quote}
Screenshots show the author using XCode's memory profiler, examining increasing memory consumption graphs, inspecting code functions with memory allocation annotations, and adding debugging statements around object creation - all indicating active memory leak troubleshooting.
\end{quote}

\subsection{Interfacing with the \model{}}
\label{sec:python_interface}
Here, we outline core functions to interface with the \model{}, highlighting how applications and users can interact with the underlying abstractions. \model{}s are implemented in Python can be pip-installed as a package. Later in \S\ref{sec:gum_tech_start}, we revisit how \model{} implements each of these functionalities.
\label{sec}

\paragraph{Instantiating a \model{}} Instantiating a \model{} requires access to an instruction-following (V)LM (e.g., GPT, Claude, Llama). We use Llama 3.3 70B running on secure servers to maintain privacy, though any model supporting the OpenAI ChatCompletions API is compatible. All propositions and associated metadata are located under the private \texttt{GUM.\_propositions} variable, accessible only to the user. Applications cannot directly manipulate the private \texttt{propositions} variable attached to the \model{}, unless explicitly granted permission by the user.

\paragraph{Subscribing to observations} The \model{} can \texttt{subscribe} to various information streams with user permission. Like traditional context-aware systems that require  operating system hooks~\cite{dragunov2005tasktracer}, \model{} also exposes a handful of default hooks, which we call Observers. These Observers listen for new observations related to screen content, notifications, and manual input. For example, when a user saves a file or moves their mouse, the Filesystem and Screen Observers pass new observations to the \model{}, which converts these observations into propositions.

\paragraph{Querying the \model{}} Users and applications alike can search for propositions (and underlying observations) using natural language queries. The query function supports parameters for search terms, response diversity, and time filtering. For example, searching for "HCI conference" will return relevant propositions about the user's HCI-related activities. The query function also supports options to control a handful of parameters: the balance between the relevance and diversity of responses (0 = maximize relevance, 1 = maximize diversity), the timestamp cutoff from which the query should start its search (looking back), and whether or not a decay should be applied based on the proposition's timestamp.
\begin{figure}
    \centering
    \includegraphics[width=0.9\linewidth]{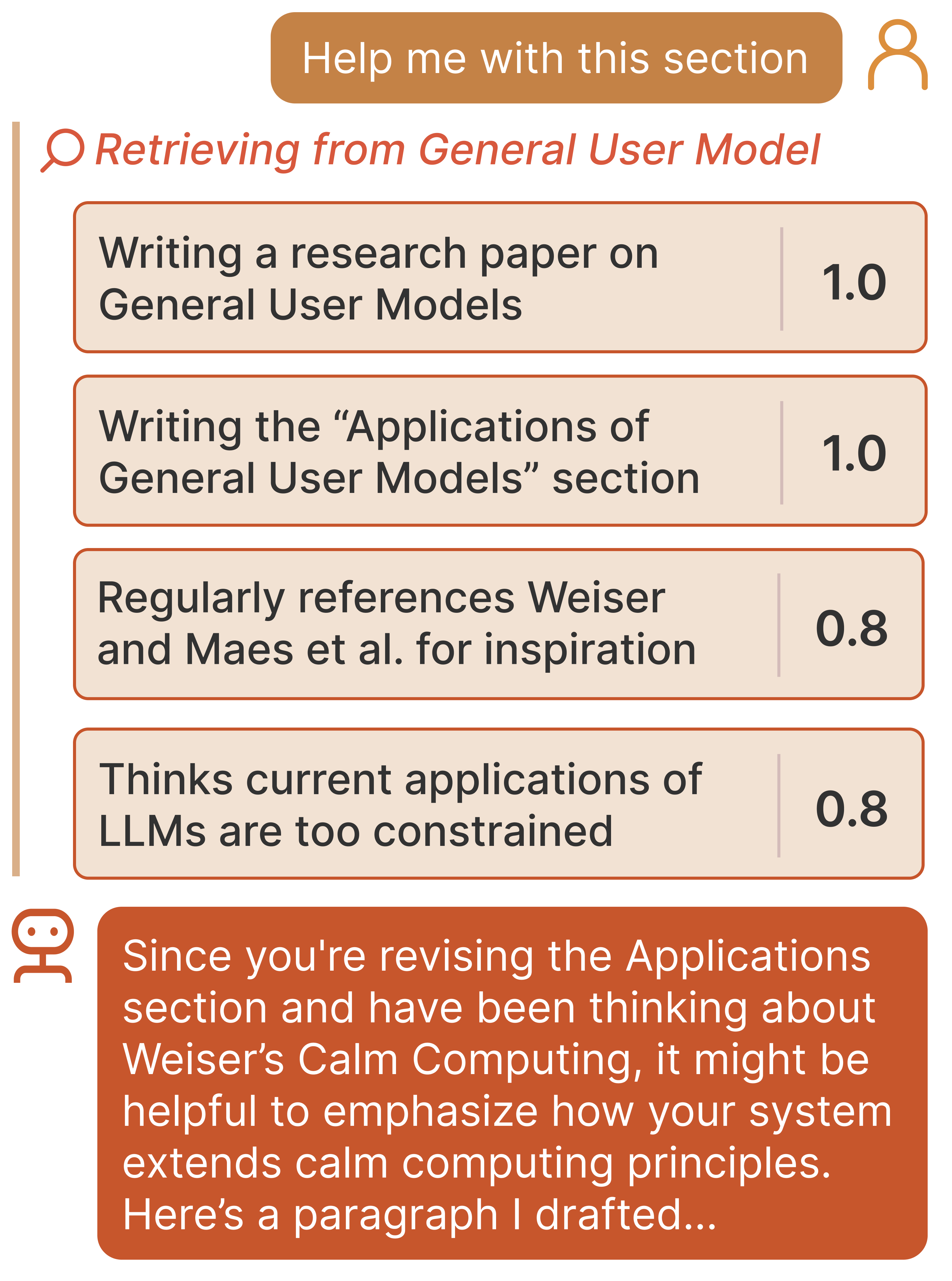}
    \caption{General User Models can ground user queries in a broader context. In this example, the user is writing an Applications section for this paper. Given a severely underspecified query, the General User Model marshalls relevant context---the draft the user has been working on, the section the user is referring to, the applications the user has already considered---and generates a highly contextualized response.}
    \label{fig:application-prompt}
\end{figure}

\section{Applications of General User Models}
To illustrate the breadth of applications for a \model{}, we outline a range of downstream use cases. These span from simple applications, such as augmenting powering chat-based LLM clients with context from \model{}, to introducing a new class of proactive systems that discover helpful suggestions for users. We also reflect on how \model{}s empower classic HCI visions, including ubiquitous computing, interface agents, and context-aware computing.

\subsection{Prompting Models}
\label{sec:prompting}
Before we get distracted by visions, let's return to something slightly more humble: prompting. A fundamental problem with prompting is a lack of shared \emph{grounding}~\cite{shaikh2025navigating} with a model. Unless explicitly instructed, LLMs have no idea what project you're working on, who you're working with, or why the prompted task was important to you. This grounding gap results in models that often incorrectly \emph{assume} aspects of our context. 
 
Prompting LLMs is inherently underspecified: we're forced to turn our rich context into a decontextualized natural language query. Here, we show how \model{} can help close the \emph{grounding gap}~\cite{shaikh2023grounding} in how we interact with chat-based LLMs. 

Consider the following prompt alongside the author's own \model{} at this point in the paper (see Fig. \ref{fig:application-prompt}).
\begin{quote}
    \textsc{User: }help me with this section.
\end{quote}
Today's frontier LLM models are lost, returning a clarification question (``Sure, I can help! Could you share the section you're referring to?''), even though there's a rich amount of context to be constructed from observation. The author's \model{} has access to interactions between the authors, papers the author has recently been reading, the state of the current paper, prior HCI work the author likes, etc. In contrast, with a standard LLM users must (at worst) painfully rebuild this context turn by turn, correcting the model when it incorrectly makes assumptions. At best, LLMs rely on context from \emph{past} chat interaction (using a sliver of the user's true context) or ask clarification questions. 

Here, the \model{} bridges this grounding gap by querying for additional propositions and their underlying observations. In our example, the \model{} returns papers the author was looking at earlier, feedback from the coauthors, and the section the author was revising.
\begin{quote}
    \textbf{Proposition}: \texttt{Omar is currently focusing on the Applications section.}\\
    \textbf{Confidence}: 0.8\\
    \\
    \textbf{Proposition}: \texttt{Omar recently read Weiser's Calm Computing.}\\
    \textbf{Confidence}: 0.6\\
    \\    \textbf{Proposition}: \texttt{Omar's advisor mentions looking into Interface Agents.}\\
    \textbf{Confidence}: 0.5
\end{quote}
Conditioned on the \model{}, LLMs generate responses that are far more personalized to the specific user.
\begin{quote}
    Since you're revising the Applications section and have been thinking about Weiser’s Calm Computing and Interface Agents, it might be helpful to emphasize how your system extends calm computing principles through contextual awareness. You could also contrast \system{}'s general assistance with the more fixed approach seen in traditional interface agents. Here’s a paragraph I drafted...
\end{quote}

\subsection{Any Application Can Query A \model{}}
Beyond prompting, any application can connect to a \model{}, using context from all \emph{other} applications and the user. Across a range of HCI systems, having access to context and being able to process it is a key single enabling layer of the application. To outline this, we revisit \citet{weiser1991computer}'s calm technology, \citet{maes1993learning}'s interface agents,  and, in a slightly ``meta'' sense, \citet{dey2001conceptual}'s context protocols.%

\paragraph{Calm Technology (\citet{weiser1996designing})} Weiser's ubiquitous computing vision launched a raft of work realizing systems that slide to and from the user's periphery. People have typically built such systems based on custom sensors \cite{fogarty2005predicting, hudson2003predicting} such as cameras and microphones, where each application has to build a custom ML pipeline to make decisions \cite{fogarty2005examining, horvitz2013lumiere}. The \model{} enables all of these applications with no such instrumentation or training set, in a way that is sensitive to the user's task. Weiser's calm technology engages both the center and periphery of a user’s attention, moving seamlessly between the two. A key design consideration for calm technologies \cite{weiser1996designing} is recognizing that what resides in the periphery can shift from moment to moment. 

For example, operating systems that use \model{}s can surface important information when it’s truly needed. For example, the first author of this paper was swamped with writing this paper, and the CHI early bird registration deadline came and went. By default, their operating system shows notifications for every email they gets in their primary inbox---causing them to generally ignore the notification, as they also received emails in the same period about ``firebase-noreply-billing'' and ``Job Opportunity at Startup''. When prompting Llama 3.3 with ``Filter out notifications that are important to me.'', it suggests passing all of these notifications through to the author, leading to attention overload. However, the same prompt, augmented with \model{}'s propositions and confidence, provides context that the author (1)~``is quite happy as a Ph.D. student'', (2)~``has a major impending deadline'', and (3)~``regularly ignores billing notifications'' As a result, the LLM now allows only the CHI registration notification through.

\paragraph{Interface Agents (\citet{maes1993learning})} 
Interface agents introduced personalized and proactive assistance for human-computer interaction. Instead of placing the burden of initiative on the user, interface agents are applications that aim to anticipate user needs and automate tasks. This requires both domain knowledge and personalized user models, but an implementation has remained challenging without a continuously learned model of individual preferences. Each new agent requires specialized datasets and hand-crafted learning objectives to capture evolving user preferences~\cite{maes1993learning}.

\model{}s enable interface agents more broadly. They supply both the user model and reflect the user's preferences across time. Consider Maes' Firefly agent~\cite{shneiderman1997direct}: a movie recommendation agent that proactively surfaces interesting new movies based on the user's preferences. For the author, now would be a horrible time---they are scrambling to finish their paper. But when the author prompted a standard LLM (Llama 3.3 70B) and asked if now was a good time for a movie, it agreed \emph{and} proceeded to recommend a list of movies that were completely unrelated to the author's interests (psychological thrillers). In contrast, with the same prompt, the \model{}-augmented LLM recognizes that the author (1) has ``many comments unresolved on his draft'' and (2) has only ``a few hours until the paper deadline''. As a demonstration, when we ablate everything deadline-related from the \model{}, the \model{}-augmented LLM correctly agrees and recommendes a relevant list of movies. With \model{}s, we can instantiate a range of proactive personalized interface-agents, with minimal domain-specific engineering and a handful of simple prompts (e.g. \texttt{Should I watch a movie?}).

\paragraph{Context-Aware Computing} Unifying context across applications is a long-standing vision of context-aware computing, but creating effective context protocols~\cite{dey2001conceptual} has remained evasive. If this were achieved, a common API could then be queried by any application to adapt based on relevant user context. Applications built using context toolkits---from context-aware reminder systems \cite{dey2000cybreminder} to assistants for conference attendees \cite{dey1999conference}---relied on infrastructure for specialized sensors. Supporting a new \emph{specialized} sensor is a challenge: for each new sensor, an entirely new data processing pipeline must be manually engineered. 

\model{}s promise to solve exactly this! Any input data type---a screenshot, sensor data, videos, etc.---can, in principle, be embedded in the \model{}. As an example, the first author exported their self-reported physical exercise logs from a weight training app on their phone and serialized the day-level logs into text files. In addition, they exported step counts from their Apple Health app. With no data preprocessing (beyond initial .txt translation), the first author fed these observations directly into their \model{}. The result:
\begin{quote}
    \textbf{Proposition:} \texttt{Omar is prioritizing their paper deadline over routine physical exercise.} \\
    \textbf{Confidence:} 0.6
\end{quote}

\begin{figure*}
    \centering
    \includegraphics[width=\linewidth]{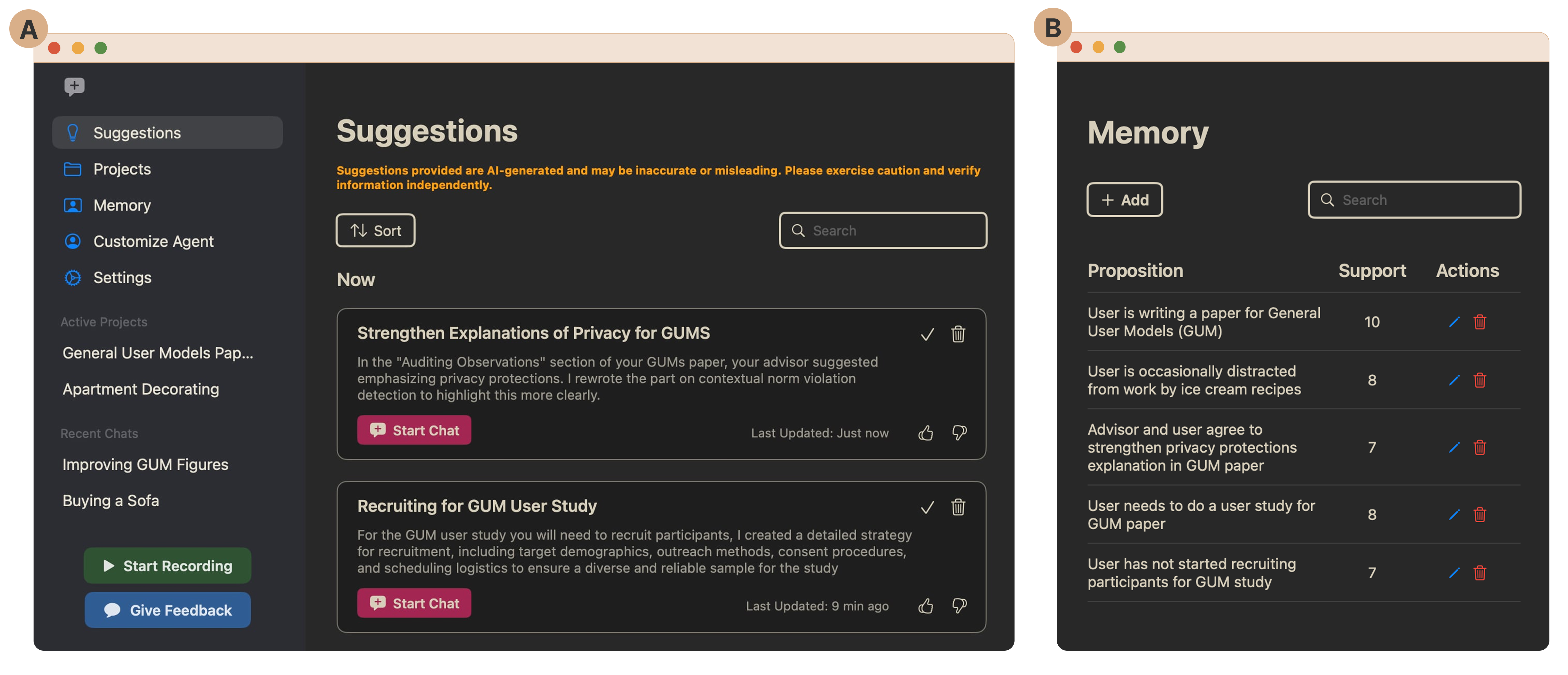}
    \caption{GUMBO is an assistant designed to proactively surface and execute helpful suggestions for users. We built \model{} into a custom desktop app. (A) The Suggestions page displays suggestions to the user, each of which the system has attempted to complete as much as possible on its own. In this figure, GUMBO suggested that the author recruit participants for the GUM user study and put together a detailed strategy plan in the background. Users can hit "Start Chat" to talk to GUMBO in more detail. (B) The Memory page allows users to view the raw propositions in their GUM, and edit, delete, or add propositions.}
    \label{fig:fig-gumbo}
\end{figure*}

\subsection{\system{}: an Assistant Built On \model{}s}

To more concretely ground the examples above, we now describe how \model{}s enable a class of applications that can proactively discover suggestions based on user context. We introduce \system{} (Fig. \ref{fig:fig-gumbo}), an assistant that ingests screenshots from a user's screen. Using screenshots, \system{} builds an internal \model{}. With the \model{}, \system{} discovers helpful suggestions, determines if a suggestion is worth showing to a user and executing, and then executes the (sandboxed) suggestion to the best of its ability---sharing preliminary results with the user.

\system{} demonstrates how to construct a system that utilizes \model{} in many different capacities simultaneously: (1)~generating rapid, relevant suggestions that the system can do on behalf of the user or help the user complete; (2)~estimating how much utility each suggestion will provide to the user if completed, enabling Horvitz's mixed initiative interaction framework~\cite{horvitz1999principles}; (3)~revising the model with explicit user feedback. 

\subsubsection{Discovering Suggestions} Before suggesting anything to a user, we must generate a candidate set to evaluate. \system{} retrieves the latest propositions from \model{} and uses them to generate candidate suggestions. It then employs \model{} to rank and filter these suggestions, presenting only a curated subset to the user.

When a new proposition is constructed, e.g., \texttt{User is likely going to friend's wedding in Chicago}, \system{} retrieves related propositions $G$ using the \texttt{query} function provided by \model{} (\texttt{User doesn't own any suitable formal wear} from recently browsing clothing sites, \texttt{User needs to budget} based on recent bank account checks, etc.). These propositions are pulled based on semantic relevance to "wedding travel" and the recency of related observations (e.g. pulling up a bank statement or starting a search for buying formal wear online). 

Using the set of related propositions $G$, along with all the underlying observations attached to each proposition, we generate a set of candidate suggestions $\tau_{i} \sim P_{LM}(* \mid G)$ (e.g. \texttt{Search for cheap suit rentals in Chicago}). To generate candidates, we prompt the underlying LLM, Llama 3.3 70B, with all the propositions and underlying metadata. An abridged prompt is below:
\begin{quote}
    \texttt{\{raw observations omitted\}}\\
    \\
    \textbf{Proposition}: \texttt{Omar is likely going to friend's wedding in Chicago.}\\
    \textbf{Confidence}: 0.8\\
    \\
    \textbf{Proposition}: \texttt{Omar doesn't own any suitable formal wear.}\\
    \textbf{Confidence}: 0.6\\
    \\    \textbf{Proposition}: \texttt{Omar needs to budget as a Ph.D. student.}\\
    \textbf{Confidence}: 0.5\\
    \\
    \texttt{What concrete suggestions do you have for the user based on the provided context?}
\end{quote}
For each suggestion, our full prompt also elicits additional confidence in the generated suggestion being something the user would find any value in. This can be interpreted as an aggregate suggestion confidence ($P(\tau_{i} \mid G)$), which will come in handy when we decide if it's worth showing a user $\tau_i$. For each new proposition, we use the prompt above to generate five suggestions together. 

\subsubsection{Determining if suggestions are worth showing} \system{} can generate and suggest a slew of suggestions that are relevant to the user's context. In many cases, suggestions are repeats---we simply filter using lexical overlap heuristics. The real challenge arises when we have many different, potentially useful, suggestions.

We apply mixed-initiative interaction~\cite{horvitz1999principles} principles to balance helpfulness against intrusiveness.  This approach requires estimating both the probability of a suggestion being useful and its utility to the user. By calculating the expected utility of interruption versus non-interruption, we can make informed decisions about when to surface suggestions. We weigh the probability the user would find some value in the suggestion $P(\tau_{i} \mid G)$ against the benefits of suggestion completion $B$ and the costs of false positives $C_{FP}$ and false negatives $C_{FN}$.

Mixed initiative interaction assumes some model that produces costs and benefits for a specific suggestion---a major cold start hurdle for the practical implementation of mixed-initiative systems to date. In our case, we turn again to the \model{}, which can estimate these quantities with no further training. Conditioned on the propositions from the \model{}, we elicit costs and benefits using a prompt (Appendix \ref{appdx:mii_prompt})---similar to how we generated suggestions. With the costs, benefits, and probabilities, we can compute the expected utility of (not) interrupting ($E[U]$) the user with the suggestion: 
\begin{align}
E[U_{\text{interrupt}}] 
&= P(\tau_{i} \mid G) \cdot B \;+\; (1 - P(\tau_{i} \mid G)) \cdot (-C_{FP}). \\
E[U_{\text{$\neg$interrupt}}] 
&= P(\tau_{i} \mid G) \cdot (-C_{FN}) \;+\; (1 - P(\tau_{i} \mid G)) \cdot 0, \\
&= P(\tau_{i} \mid G) \cdot (-C_{FN}).
\end{align}
To determine a decision threshold, we can simply test if the expected utility of interrupting is greater than not interrupting.
\begin{align}
E[U_{\text{interrupt}}] > E[U_{\text{$\neg$interrupt}}]
\end{align}
Occasionally, we found that suggestions would still pour through the decision boundary. The issue arises because \citet{horvitz1999principles}'s mixed-initiative framework treats actions as independent---here though, the utility of adding an additional suggestion depends on how many other suggestions are being presented to the user in the same time period. To address this, we implement an additional token-bucketing algorithim~\cite{demers1989analysis}, rate-limiting suggested suggestions to a maximum of 1 per minute. Altogether, we can re-use \model{} to determine when to interrupt in the first place, operationalizing principles from mixed initiative interaction.

\subsubsection{Ideas are cheap. Execution is everything}
Instead of solely suggesting that the user \texttt{Find a suitable place to rent a suit in Chicago}, \system{} should go a step further and search for rental locations on its own and report its findings. In other words, if \system{} is capable of completing the suggestion---and doing so does not cause irreversible side effects (such as actually ordering a suit)---it should execute the suggestion itself and present the results to the user.  

When necessary, suggestions generated by \system{} can be delegated to a set of tools or agents. We use an additional zero-shot prompt to determine if a suggestion requires a tool (see Appendix \ref{appdx:tool_needed}). We implement a handful of external tools which can be toggled by the user (and are by default disabled) to preserve privacy. \system{} delegates calls to Gemini 2.0 Flash \emph{only} if search with Google or code execution is required. In addition, \system{} implements file system search and retrieval through the local MacOS spotlight search, and computer use via the Operator API.\footnote{We leave Computer Use disabled in \system{}'s evaluation, both to reduce external dependencies and since we found computer use too slow / buggy. We expect advancements in computer use to greatly benefit from implementing \model{}s.}

\subsubsection{Feedback.} Finally, \system{} allows users to leave thumbs up / down or natural language feedback on any suggestion. \system{} simply converts the feedback into a text representation (\texttt{User disliked the following suggestion: [suggestion]}) and feeds it back into \model{}. We treat feedback the same as any other unstructured observation, placing the onus on the \model{} to convert feedback into appropriately weighted propositions. 

\begin{figure*}
    \centering
    \includegraphics[width=\textwidth]{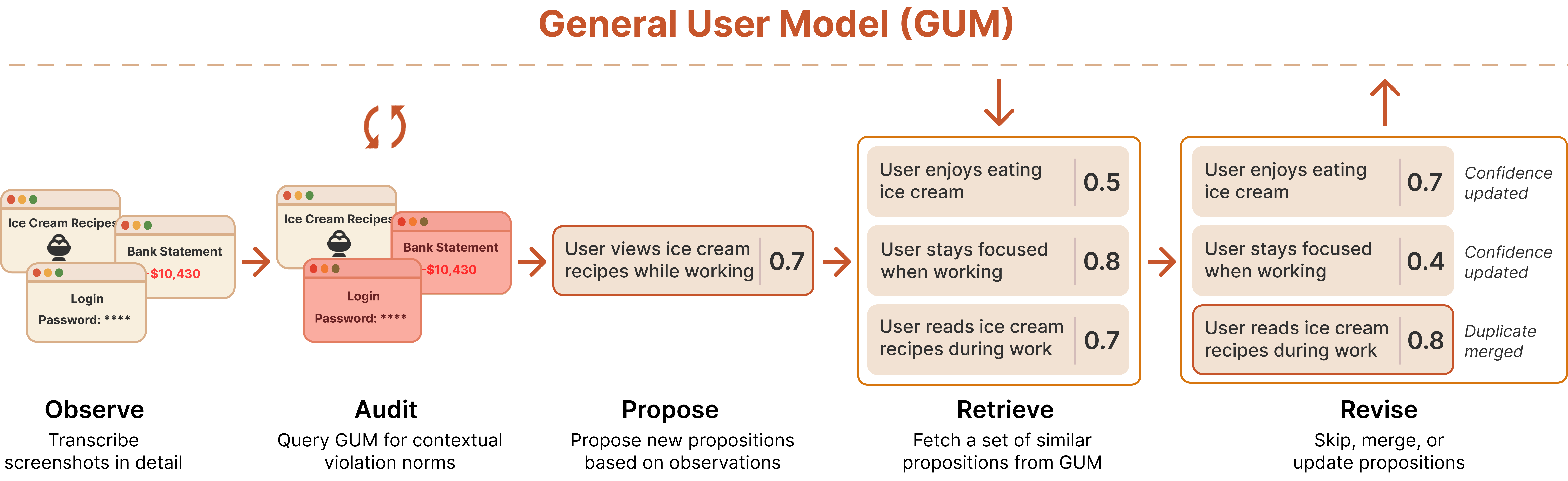}
    \caption{The GUM pipeline observes user behavior via unstructured inputs (e.g., screenshots), audits updates for privacy violations using contextual integrity, constructs propositions with associated confidence scores, retrieves contextually similar propositions, and revises the model. Each step iteratively refines GUM’s understanding of the user.}
    \label{fig:enter-label}
\end{figure*}

\section{Constructing a General User Model}
\label{sec:gum_tech_start}
To continuously learn a user model, \model{}s \textit{construct}, \emph{retrieve} and \textit{revise} propositions about a user by ingesting completely unstructured \emph{observations}. In this section, we describe the architecture behind these components, and explain how they produce the GUM API.

Our general engineering principle here is to rely primarily on open-source models. While closed-source models are more performant~\cite{bommasani2021opportunities}, we expect \model{}s to be \emph{owned} by individual users and eventually distilled to be run on their local devices. By focusing on open-source models, we demonstrate and advocate for the core \model{}s to operate without sending private user data to third party AI platforms. As gaps between closed and open sourced models close and as models become cheaper for inference, \model{}s will become more performant and feasible on commodity hardware.

Powering the \model{} in this paper are two models. To observe screen interactions, we use Qwen 2.5 VL~\cite{bai2025qwen2}---a vision-language model with a vision encoder. For proposing/revising propositions, we use Llama 3.3 70B~\cite{grattafiori2024llama}, a language-only decoder model.  

\subsection{Observing Interaction}
The basic substrate of the \model{} is user behavior. Where prior user models were often built off narrow, manually-logged observations (e.g.,  \citet{fischer2001user},  \citet{jameson2001modelling}, \citet{horvitz2013lumiere}), the inference abilities of modern VLMs open up opportunities for unstructured, unobtrusive inputs such as screenshots, files, or streams of open-ended text as the \model{}'s raw observations. The \model{} Python package (introduced in \S\ref{sec:python_interface}) provides a handful of hooks that return raw observations. These hooks---Observers---are generic by design. For example, an audio Observer might connect to the device's microphone, transcribe the user's speech, and return that speech as a text string to the \model{}---nothing else.

We implement a handful of Observer instantiations in Python for MacOS. The \texttt{\textbf{Screen}} Observer listens to I/O activity (keyboard, mouse clicks), captures a screenshot of the screen on I/O input, and transcribes the contents of the changes into a text update for the \model{}. It relies on a VLM (Qwen 2.5 VL 72B) to convert screenshots into a transcript. Beyond the content, \texttt{Screen} also generates a description of actions the user takes across up to 10 unique frames (example outputs in Appendix \ref{appdx:screen_outputs}). A \texttt{\textbf{Notification}} Observer watches for updates on the operating system's underlying notification database and returns  updates that contain the notification text and contents. Our provided Observers---\texttt{Screen} and  \texttt{Notification}---are just examples; one could implement new Observers for any modality, e.g. audio input, health sensor data, videos.

\subsection{Auditing Observations}
\label{sec:audit}
Even though \model{} uses open-source vision-language models deployable locally, users may accidentally expose private information they never intended for processing. To prevent this, we implement an \emph{Audit} module that filters listener updates before they reach the final collection of propositions.

\model{}s learn a user's privacy expectations by observing data-sharing norms, using itself to infer user-specific models of contextual integrity. Consider two updates: (1) a user writing a work email about a conference versus (2) entering bank credentials on a financial website. The Audit module would allow the first as typical professional communication while blocking the second as sensitive financial data. In developing the Audit module, we instantiate \citet{nissenbaum2004privacy}'s contextual integrity theory, which preserves privacy when information flows match social context norms, focusing on appropriate sharing rather than binary data control.

Imagine a computer science professor's \model{}. The \model{} has already observed the user discussing course materials with colleagues via email and sharing research papers. However, it has never observed the user sharing login credentials or financial information with anyone. When the user enters their banking password on a financial website, the Audit module uses \model{} to retrieve past propositions related to the current observation. These include:
\begin{quote}
\texttt{Michael is careful about sharing access credentials} \\

\texttt{Michael has discussed privacy concerns with students in the past} \\

\texttt{Michael only discusses financial matters with his spouse} \end{quote}
Using these retrieved propositions, the Audit module then generates answers to \citet{nissenbaum2004privacy}'s contextual integrity questions:
\begin{quote}
\texttt{1. Is the user disclosing any new information?}\ \texttt{\textbf{Yes}}
\\
\\
\texttt{2. What type of data is the user disclosing?} \ \texttt{\textbf{Banking credentials and financial account information.}}
\\
\\
\texttt{3. Who is the primary subject of the data?} \ \texttt{[User's name]}
\\
\\
4. Who is the recipient? \\ \texttt{An AI model that ... [description of GUM]}
\\
\\
5. Should this data be transmitted to the model?\\ \texttt{\textbf{No}}
\end{quote}
The Audit module determines this observation contains sensitive financial information that the user would not want shared, and blocks the update from progressing to the remaining pipeline. 

\subsection{Constructing Propositions}
\label{sec:how_prop}
We now turn to making inferences about the user from observation. The Propose module converts observations into a set of propositions. During this process, Propose generates a reasoning trace that ties observations to a proposition and produces a confidence score associated with each proposition. 

\paragraph{Reasoning About and Generating Propositions} Before generating a proposition, we want to ensure that the model reasons over \emph{how} the observations are connected to the proposition. This rationale serves two purposes: it gives users some explanation for why an inference was made and it generally improves model performance~\cite{nye2021show,wei2022chain}. In this initial reasoning step, we first instruct the model to generate an explanation related to the observation's relationship to the user.
\begin{quote}
\texttt{observation: [screenshots of the user switching between Overleaf and YouTube]}
\end{quote}
With this observation, the \emph{Propose} step generates the bolded \textbf{rationale}, explaining the user's behavior.
\begin{quote}
\texttt{proposition\_reasoning: \textbf{``The user appears distracted, switching focus between an ice cream recipe video and typing intermittently in an Overleaf window.''}}  
\end{quote}
Once the rationale is generated, both the observation and rationale are jointly used to generate a final proposition, e.g.  $\mathrm{proposition}\sim P_{LM}(\cdot \mid \mathrm{reasoning},  \mathrm{obs})$
\begin{quote}
    \texttt{proposition: \textbf{``User periodically views ice cream recipes while writing.''}}   
\end{quote}
This is true---the first author is an ice cream afficionado, and their preferred writing distraction is browsing ice cream recipes.

\paragraph{Producing Confidence Scores Over Propositions} Once the proposition and rationale are generated, we use \model{}'s underlying LLM to produce a confidence score in the proposition. Confidence scores support future revisions of uncertain propositions and help applications make decisions under uncertainty. While we could, in theory, use the LLM's own log probabilities by looking directly at the logit scores on the completion, out of the box model predictions are often overly confident. Like~\citet{tian2023just}, we observe that simply prompting the model to generate confidence scores yields more calibrated outputs instead of looking at underlying logits.\footnote{Our setting is slightly different: we're looking at confidence in a user's beliefs, instead of confidence in the model's own response---but \citet{tian2023just} generalizes.} We prompt the model to generate a confidence score on a 1-10 scale, then normalize between 0-1. Our full prompt is in Appendix \ref{appdx:calibration}.

\paragraph{Proposition-specific decay} Some propositions remain more relevant over time compared to others. Unfortunately for the first author, a proposition like \texttt{``User is eating ice cream.''} decays faster than \texttt{``User is Ph.D. student.''} (Such is the nature of life.
But, a proposition such as \texttt{``User enjoys eating ice cream more than their Ph.D.''}\footnote{Yes, a real proposition generated by our system.}---that’s timeless.) To account for this, we take inspiration from cognitive architectures~\cite{anderson1993rules, park2023generative} and introduce a decay score. Instead of applying a global decay to all propositions universally, we use the underlying LLM to generate a proposition-specific decay. We include all information about the proposition in-context (reasoning and confidence) and generate a decay score from 1-10, e.g. $\mathrm{decay} \sim P_{LM}(\cdot \mid \mathrm{confidence}, 
\mathrm{proposition},
\mathrm{reasoning}, \mathrm{obs)}$.

\subsection{Retrieving Context-Sensitive Propositions} 

The process so far produces propositions and confidence scores (from Propose, \S \ref{sec:how_prop}), but we need to compare these with what \model{} already knows. To enable comparison and retrieval over \model{}s, we implement a \emph{Retrieve} module. Speed is critical: each observation generates multiple propositions that require hundreds of queries. We use a two-stage \textbf{retrieve} and \textbf{rerank} approach~\cite{nogueira2019passage}. First, we \textbf{retrieve} propositions and underlying observations (our ``document'') using BM25, which considers term frequency, inverse document frequency, and document length.\footnote{While we use lexical similarity, one can easily replace our approach with neural embeddings} Then, we use an LLM-based \textbf{re-ranker} to improve precision. Our setup optimizes for speed but can accommodate other retrieval/reranking methods.

\paragraph{Recency-Adjusted Relevance}
BM25 computes relevance scores, but recent propositions should be prioritized. Each proposition $d_i$ includes a decay score $\alpha_{i}$ (\S\ref{sec:how_prop}). We define the decay score as $\gamma_i = \exp\bigl(-\alpha_{i} \cdot k \cdot \text{age}(d_i)\bigr)$, where $k=2$ and $\text{age}(d_i)$ is a continuous value measured in days. The adjusted relevance score is:
\[
\tilde{r}_i = r_i \cdot \gamma_i.
\]
\paragraph{Incorporating Diversity via Document Similarity}
Retrieved propositions should be diverse. We use Maximum Marginal Relevance (MMR) to balance relevance and diversity~\cite{carbonell1998use}. Using TF-IDF vectors for document representation, we compute cosine similarity between documents. Given a set $S$ of selected propositions, the diversity term for a candidate document $d_i$ is:
\[
\delta_i = \begin{cases}
\displaystyle \max_{d_j \in S} \text{sim}(d_i, d_j) & \text{if } S \neq \emptyset
\end{cases}
\]
The MMR score combines relevance and diversity:
\[
\text{MMR}(d_i, S) = \lambda\, \tilde{r}_i - (1-\lambda)\, \delta_i,
\]
where $\lambda=0.5$ balances the trade-off. We iteratively select documents maximizing MMR until reaching $N$ documents.

\paragraph{Reranking and Filtering.} We pass retrieved propositions through an LLM re-ranker that classifies them as \emph{identical}, \emph{similar}, or \emph{unrelated} (the reranker prompt is in the Appendix \ref{appdx:reranker}). For example, when querying with \texttt{User is periodically distracted by ice cream recipes}, a proposition like \texttt{User is a CS Ph.D. student} would be labeled as unrelated. On the other hand, \texttt{User might have a sweet tooth} would be related, while \texttt{User appears distracted by ice cream recipes} would be identical. Successive identical propositions might signal that a revision is necessary, which we address next in \S\ref{sec:how_rev}.

\subsection{Revising Propositions}
\label{sec:how_rev}
As the breadth and quantity of propositions in the \model{} grows, revision becomes critical. The retrieve module simply surfaces similar propositions. However, newly generated propositions in \model{}---flagged as \emph{similar} by our retrieval module---can contradict or reinforce prior propositions. The final \model{} should reflect these differences. We introduce a \emph{Revision} module that takes newly generated propositions and appropriately revises prior ones. Over time, the revise step refines the \model{}, yielding a collection of propositions that reflect a richer understanding of the user. 

To implement revision, we construct a prompt that takes as input retrieved propositions and the newly generated propositions, along with the underlying observations for each proposition. The revision process can operate over multiple input/output propositions, revising multiple propositions at once. Using the newly generated propositions, the revision module rewrites the proposition, regenerates confidence, and then revises associated metadata. We encode both the past proposition(s) and the new proposition(s) in a prompt. We additionally include all metadata associated with both the old and new propositions---the underlying grounding, reasoning traces, and decay scores.
\begin{quote}
    \texttt{\#\# Past Proposition}
    \\ 
    \texttt{Proposition: User periodically views ice cream recipes while writing.}
    \\
    \texttt{Confidence: 6}
    \\
    \texttt{Decay: 4}
    \\
    \texttt{[additional metadata (grounding, reasoning)]} \\
    \\
    \texttt{\#\# New Proposition}
    \\
    \texttt{Proposition: User messaged ice cream recipes to colleague during a meeting.}
    \\
    \texttt{Confidence: 6}
    \\
    \texttt{Decay: 2}
    \\
    \texttt{[additional metadata (grounding, reasoning)]}
\end{quote}
Conditioned on both the past and current proposition, we also generate a revised proposition: $\mathrm{revision}\sim P_{LM}(\cdot \mid \mathrm{prop_{new}}, \mathrm{prop_{old}})$.     Generated portions are \textbf{bolded}.  
\begin{quote}
    \texttt{\#\# Revised Proposition}
    \\
    \texttt{[regenerated reasoning]}\\
    \texttt{Proposition: \textbf{User is \emph{regularly} distracted by ice cream recipes}}
    \\
    \texttt{Confidence: \textbf{10}}
    \\
    \texttt{Decay: \textbf{1}}
\end{quote}
In the example above, our revised confidence increased. However, revision can also yield propositions with reduced confidence, especially when contradictory propositions are compared. We never completely evict propositions that are revised to zero confidence: they remain in the \model{} for transparency. However, queries to the \model{} will---by default---not surface confidence = 0 propositions.

\section{Evaluating General User 
Models:\\Accuracy and Calibration}
\label{sec:tech_eval}

\model{}s aim to accurately capture a user's context from unstructured observation. To first evaluate whether \model{}s can effectively synthesize context into accurate inferences about users, we conduct an evaluation in a controlled setting using email data. We assess \model{}'s \emph{accuracy} compared to ablations and evaluate \model{} \emph{calibration}---whether confidence levels appropriately reflect uncertainty.

We chose email as our evaluation domain because inboxes contain diverse and relevant information across modalities, while also being full of irrelevant content. Email overload makes synthesizing important information difficult: \model{}s must maintain accuracy despite this challenge. Though email represents only a subset of user interactions, its controlled, semantically rich nature provides a suitable testbed for annotation.

\subsection{\model{} Ablations} 
\label{sec:tech_ablations}
For a controlled evaluation, our hypotheses require baselines for evaluation. Generating an accurate proposition can be trivial---\model{}s could generate obvious and uninteresting facts about the user (e.g. \texttt{[ANON] is using their computer}). We therefore evaluate relative quality between generated propositions, using the following ablations as conditions:
\begin{itemize}
    \item \textbf{\textsc{Ablation:} \texttt{No \{Retrieve, Revise\}}}. In this ablation, we test a version of \model{} that only \emph{creates} propositions with no confidence. The \model{} is unable to retrieve past context---propositions are constructed given only the current email. 

    \item \textbf{\textsc{Ablation:} \texttt{No \{Retrieve\}}}. Here, we enable just the Revision model, revising old propositions based on current emails using a sliding window of past propositions (that fit into the model's context window) without the full retrieval engine.
    
    \item \textbf{\textsc{Full:}} The full \model{} architecture uses both the Revise and Retrieve modules. enabling retrieval over longer timespans and appropriate proposition revision.
\end{itemize}

\subsection{Procedure}
Our procedure involves a data loading process, where the participant exports the last 200 emails (threads, attachments, links, etc.) onto their personal computer. Then, participants train their own \model{}, sequentially feeding inputs in the same order they appear in the inbox. Finally, we ask participants to label and rank outputs from the \model{} and its ablations for quantitative evaluation, and share open-ended feedback on aspects of the study. 

\subsubsection{Participants and Infrastructure} 
\label{sec:tech_participants}
\paragraph{Participants} We recruited $N=18$ participants through a mix of snowball sampling, word-of-mouth, and workplace channels at our institution. We required participants to be over 18 years old, be fluent in English, and have an active Gmail account. The median age was 26, with 8 identifying as male and 10 as female. 39\% identified as Caucasian, 50\% identified as Asian, and 5.5\% as Hispanic, and 5.5\% as other. Most of our participants were recruited from an academic institution: 14 of our participants have a bachelor's degree, 3 have a master's degree, and 1 has a GED.  

\paragraph{Infrastructure} In serving \model{}, we \emph{only} use open-source models, relying on Llama 3.3 (70B parameters) for proposition and revision steps. We used quantized versions for inference speed and memory efficiency on 80GB NVIDIA H100 GPUs (which were provisioned by us directly and only accessible to the research team), using the Skypilot~\cite{yang2023skypilot} library for autoscaling and load-balancing.

\subsubsection{Procedure Details} 

Participants first exported 200 emails from their Gmail "Primary Inbox" (excluding Promotions and Social emails) via an export script that preserved read/unread states and replies. They then executed a training script that sequentially fed these emails into \model{} in chronological order, with processing occurring on our private infrastructure while propositions and metadata were saved locally. Following training, each participant evaluated 30 stratified propositions across both time and confidence (ten from each condition) by providing binary accuracy judgments and ranking sets in pairwise comparisons with allowance for ties. Finally, participants completed a survey with open-ended questions (in Appendix \ref{appdx:survey}) about the strengths and weaknesses of the generated propositions, providing qualitative insights to complement the quantitative evaluations.

\subsection{Evaluation Methods}
\label{sec:eval_specific}
To evaluate \model{}, we assessed accuracy, calibration, and relative performance across ablations. We computed accuracy across the labeled propositions and compared results via t-tests. For relative rankings, we converted participant comparisons into pairwise win rates with confidence intervals, and applied significance testing using Holm-corrected binomial tests. To measure calibration, we calculated Brier scores (mean squared difference between predicted probabilities and actual outcomes), with lower scores indicating better calibration.
\begin{equation}
\text{Brier} = \frac{1}{n} \sum_{i=1}^{n} (p_i - y_i)^2
\end{equation}
Because \model{} generates confidence scores on a $1-10$ scale, we normalize these scores to the [0,1] range before computing the Brier score. Finally, we conducted t-tests for Brier across \model{} ablations.

\begin{figure}
    \centering
    \includegraphics[width=\linewidth]{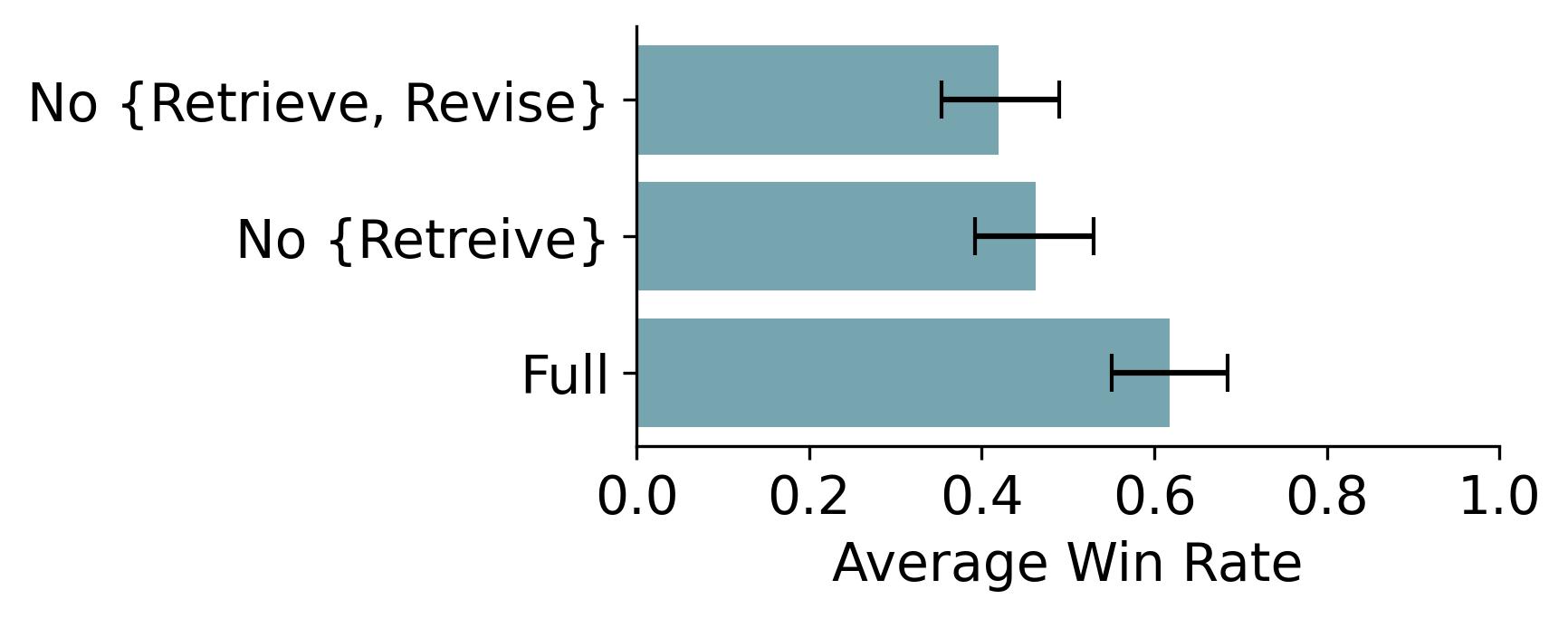}
    \caption{Relative to ablations, propositions from the full \model{} architecture are preferred by users. Additionally, retrieval and revision depend heavily on each other. Without retrieving relevant propositions, the revise module is unable to correctly adjust propositions. Ablating retrieval causes win rates to fall significantly.}
    \label{fig:ablation_ranking}
\end{figure}

\begin{table}[]
    \centering
    \begin{tabular}{l|rr}
    \toprule
    Variants & Accuracy & Calibration (Brier) \\
    \midrule
     No \{Retrieve, Revise\} & $74.41 \pm 7.81$  & --\\
     No \{Retrieve\}    & $73.28 \pm 4.42$  & $0.36 \pm 0.07$\\
     Full    & $76.15 \pm 5.85$ & $0.17 \pm 0.03$  \\
     \bottomrule
    \end{tabular}
    \caption{In isolation, accuracy for \model{} and its ablations is about the same---models generally construct correct inferences about the user. However, the full \model{} architecture produces propositions that are significantly more calibrated.}
    \label{tab:acc_res}
\end{table}

\begin{figure}
    \centering
    \includegraphics[width=0.8\linewidth]{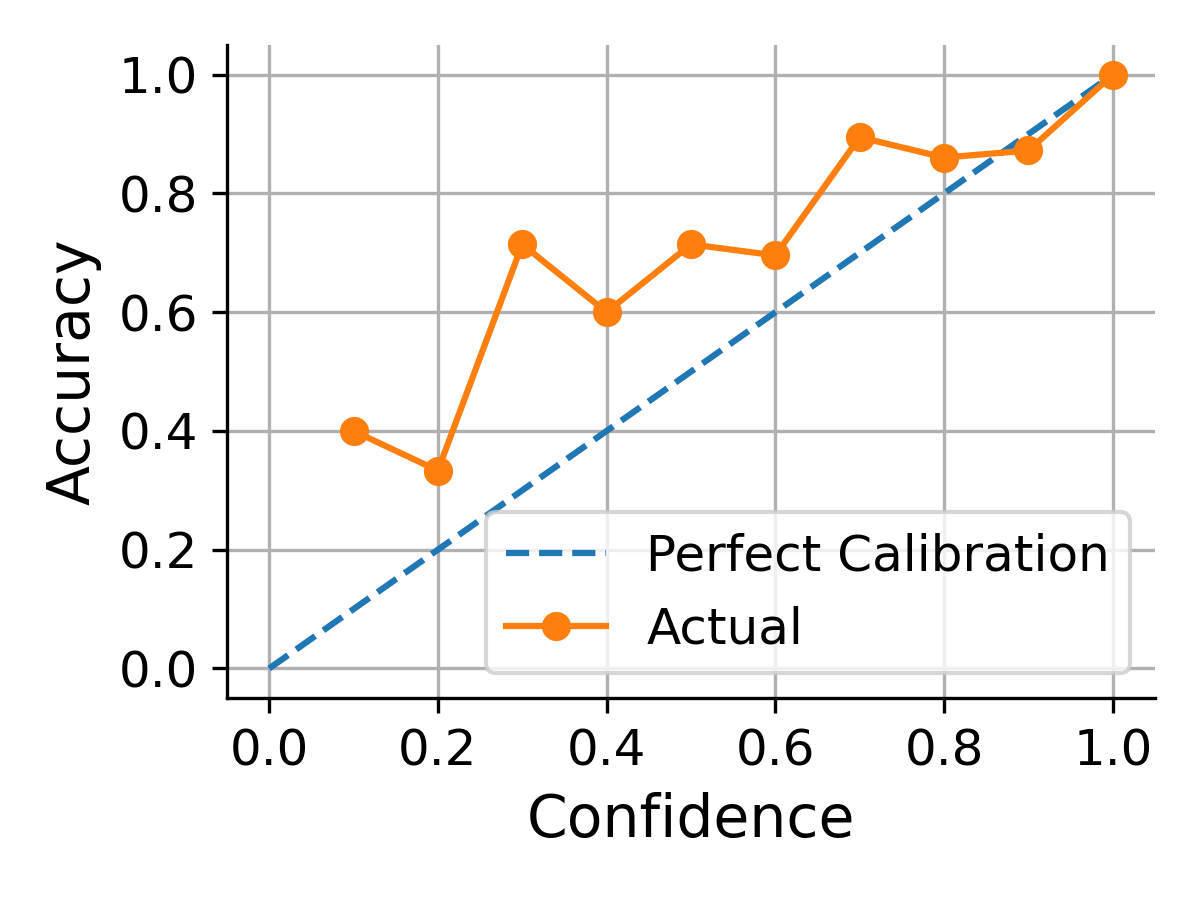}
    \caption{\model{}s are generally well calibrated. When errors occur, \model{}s are \emph{underconfident} in their propositions---the actual model's predictions lie above perfect calibration. In the user modeling setting, this is ideal. We should underestimate propositions to avoid eroding user trust. }
    \label{fig:calibration}
\end{figure}

\subsection{Results}
We outline a sample of anonymized and correctly annotated propositions from our evaluation below to illustrate the breadth of inferences constructed from email alone.
\begin{quote}
    \texttt{[ANON] values their privacy and is interested in how their personal information is used.} \\

    \texttt{[ANON] is likely a frequent traveler, given their engagement with emails about airline transfer bonuses and credit card rewards.} \\

    \texttt{[ANON] is annoyed by notifications from Academia.edu.} \\

    \texttt{[ANON] is proactive about managing their notifications and email subscriptions.} \\

    \texttt{[ANON] is aware of the risks associated with investing, including the possible loss of money.}
\end{quote}

\paragraph{The full architecture bests \model{} ablations.} To differentiate between ablations and to identify relative performance of the full \model{}, we asked participants to \emph{rank} output propositions from each ablation. In the ranked setting, the difference between the full \model{} and the ablations are clear (Fig \ref{fig:ablation_ranking}). Win rates for the full model are higher than the ablations ($\mu$, 95\% conf; \textsc{Full} = $0.618 \pm 0.07$ , No \{Retrieve\} $=0.463 \pm 0.07$, No \{Retrieve, Revise\} $=0.420 \pm 0.07$). This difference is also validated by a bootstrapped significance test following the Holm-Beniforri correction ($p < 0.05$). Between the No \{Retrieve, Revise\} and No \{Revise\} ablations, however, we see no significant difference. We suspect that relevant email is often spaced apart, and that revision \emph{without} retrieval (e.g. the sliding window approach) fails to capture long-range dependencies between emails. Without this, Revision falls apart entirely.

\paragraph{Propositions generated by the user model are both calibrated and accurate.} \rev{Highly confident (confidence $> 0.8$) propositions were labeled as correct $88.2\%$ of the time by participants. When confidence $=1.0$, propositions are completely correct ($100\%$ accuracy).} \rev{Even when we include all propositions, regardless of confidence score}, participants rank \model{} propositions as correct $76.15 \pm 5.85$ of the time\rev{---lower, but still fairly accurate.} Participants were also fairly positive on the general accuracy of propositions.
\begin{quote}
    ``People should use this instead of asking for zodiac signs or MBTIs.''
\end{quote}
On average, participants also rated propositions from email alone as accurate to very accurate ($\mu =6.47$ on a 1-7 Likert scale). In addition, we find that \model{}s are well calibrated, with a Brier score of $0.17 \pm 0.03$. Almost all of \model{}s' calibration error comes from being \emph{underconfident} (see Fig \ref{fig:calibration}). In our setting, we'd prefer that \model{}s are underconfident rather than overconfident: an overconfident user model might wrongly assume it knows a user’s preferences, while underconfidence allows \model{} to adapt more cautiously.  However, generating an accurate proposition alone is too trivial a task. All \model{} ablations, for example, generate equally accurate propositions ($\mu$, 95\% conf.; \textsc{Full} =  $76.15 \pm 5.85$, No \{Retrieve\} = $73.28 \pm 4.42$, No \{Retrieve, Revise\} = $74.41 \pm 7.81$)). No condition was significantly stronger than another (see Table \ref{tab:acc_res}).

\paragraph{``It feels kind of weird to see this in writing.''} Propositions generated by the \model{} are somewhat blunt. One participant occasionally used delivery services to order food. From the sequence of emails delivered to the participant's inbox, \model{} concluded that the participant \texttt{valued convenience and was willing to pay for it}. For another participant, \model{} correctly inferred that the user \texttt{may not always have the time to read all their emails immediately.} Propositions that reveal uncomfortably accurate insights often have lower confidence scores. Still:
\begin{quote}
    ``It is not wrong per se, but it feels kind of weird seeing this in writing.''
\end{quote}

\paragraph{Emails are one-dimensional proxy for true user context.} Despite the overall accuracy of \model{}'s propositions, a few participants noted that emails were still a highly constrained representation of their context:
\begin{quote}
    ``The emails I get on this account are not representative of my life.''
\end{quote}
Errors in propositions often emerged from this fact. Some \model{}s would over-index on a particular facet of a participant's life, focusing mostly on financial decisions, a particular hobby, or a recent purchase. In our end-to-end evaluation to come (\S\ref{sec:end_to_end}), we expand the \model{}'s viewpoint substantially, feeding it screenshots from participants' computers over the course of 5 days.

\begin{table*}
  \centering
  \begin{tabular}{@{}llp{5cm}p{6cm}@{}}
    \toprule
    \textbf{ID} & \textbf{Context} & \textbf{Role / Program} & \textbf{Representative Computer Activities} \\ \midrule
    P1 & Work \& Personal & HCI graduate student & Reading news; writing a paper; communicating with friends and family \\
    P2 & Work \& Personal & HCI graduate student & Brainstorming; personal communication \\
    P3 & Work & Civil engineering graduate student & Brainstorming; qualifying-exam preparation; reading related work \\
    P4 & Personal & Data scientist & Watching NCAA games; wedding planning; looking for a new job \\
    P5 & Work \& Personal & NLP graduate student & Data processing; gaming (Twitch streams) \\ \bottomrule
  \end{tabular}
    \caption{\rev{Overview of Participant Roles and Computer Activities}}
    \label{tab:participants}

\end{table*}

\section{Evaluating General User Models:\\Privacy Audit Module}
As we expand context for \model{}, auditing observations for privacy becomes increasingly critical (\S\ref{sec:audit}). During an Audit, the \model{} uses itself to make inferences about an individual's contextual privacy preferences when processing interaction data. We explicitly test this, hypothesizing that \model{} helps when making inferences related to a user's privacy preferences. We evaluate the Audit module using the same email setting from our Accuracy and Calibration eval (\S\ref{sec:tech_eval}), with the same set of $N=18$ participants (\S\ref{sec:tech_participants}). 

\subsection{Procedure}
We asked participants to skip annotating any propositions that violated their personal privacy preferences, effectively flagging them in the process. Using this signal, we computed the number of flagged propositions overall. Participants were also presented with the \model{}'s responses to \citet{nissenbaum2004privacy}'s contextual integrity questions---these questions were used to power the Audit module in \S\ref{sec:audit}. Participants were asked to select the best generation across \model{} ablations and score the accuracy of the best \emph{Audit} response using a 1-7 Likert score. As reference, the author's anonymized contextual integrity generation is below:
\begin{quote}
Based on the emails, you have shared your:
\\ \\
1. Email address with Calendly, Nextdoor, OpenAI, Instacart, and GitHub.

2. GitHub account with [ANON], who has invited you to collaborate on a repository...
\\ \\
You have shared this data in the following contexts:
\\ \\
1. Scheduling events and meetings (Calendly)

2. Receiving recommendations and updates from your community (Nextdoor)
...
\\ \\
You have not responded to or engaged with the Nextdoor email, which could suggest that you are not interested in sharing data or interacting with that platform.
\end{quote}

\subsection{Results}
\paragraph{\model{}s generally respect contextual integrity.} From the 180 propositions annotated by participants for the full \model{}, we found only 7 propositions that were flagged as contextual integrity violations. On average, participants agreed or strongly agreed with responses to contextual integrity generations ($\mu =6.06$ on a 1-7 Likert scale). Furthermore, a majority of participants 61.1\% selected the full \model{} as generating the best contextual integrity response, compared to ablations without retrieval (22.2\%) and revision (16.6\%). As a sanity check, we confirmed that no propositions included verbatim references to credit card numbers, phone numbers, or addresses. 

\paragraph{But when they don't, it's bad} Seven observations \emph{still} bypassed the Audit step---a non-negligible number. We looked into each participant's open-ended responses for insight. One participant (two violations) mentioned some surprise at the \model{} being able to recall specific names from their email:
\begin{quote}
    ``The ones which brought up messages or social media with other people and then explicitly named some of my friends were a little surprising, though I can imagine where all that information is sourced from within my emails.''
\end{quote}
The broader conclusion: \model{}s are able to extract a surprising number of inferences from just 200 emails. For some users, we suspect that making explicit social inferences about others is a boundary for \model{}. In principle, the user can explicitly dictate their desired contextual privacy norms to their \model{}, but there is a delicate trust balance to walk. We revisit this in our discussion.

Another participant (with 3 violations) left the following response:
\begin{quote}
    ``[The proposition was] based on a phishing email I received rather than my actual email, and it had a decently high confidence rating, but it wasn't accurate at all, and so the model thought it was me and my interest rather than just a phishing attempt.''
\end{quote}
\model{}s can be maliciously re-written by advertisements and spam, opening them up to attackers. Prompt-injection attacks are a well-known VLM vulnerability~\cite{zhang2024attacking}; \emph{however}, \model{} exposes a larger surface area to exploit this attack. Just like prompt injection attacks can be mitigated by red-teaming a model, we believe that GUM injection attacks can be mitigated---though not entirely erased without further research.

\section{Evaluating General User Models:\\End-to-End via \system{}}
\label{sec:end_to_end}
In our final evaluation, we assess the overall effectiveness of \model{} through a formative study with \system{}, a system that constructs a \model{} through screenshots of the user's screen and marshalls it to generate suggestions.

\subsection{Procedure}
Our study consists of two main parts: a \textbf{burn in}, where participants configure \system{} on their computer and have it silently learn and construct their GUM; and \textbf{active use}, where participants interact with \system{} suggestions. In this section,  we outline the infrastructure for serving \model{} and detail each evaluation stage. Our study procedure and participant recruitment was approved by our institution's IRB (up to 5 participants to mitigate any risks associated with \system{} errors in a larger sample).

\subsubsection{Recruitment and Infrastructure} We use the same infrastructure in our technical evaluation, except that we deploy a vision model (Qwen 2.5 VL 72B~\cite{bai2025qwen2}) for the Screen Observer, transcribing observations from the screen. Participants were recruited using a mix of snowball sampling and word-of-mouth. Given the level of trust required to use a system like \system{}, participants were either acquaintances of the authors or acquaintances of acquaintances. Participation required about an hour for synchronous onboarding and offboarding; compensation was \$75 USD.

\subsubsection{Participant Details} No participants had any knowledge of the system before the study. Participants were located in the United States and were over 18 years of age. 2 participant identified as White, 2 as Asian, and 1 as African American, and all had bachelor's degrees. \rev{All participants reported what they primarily used their computers for (work or personal) along with a high level list of activities (see Table ~\ref{tab:participants})}.

\subsubsection{Procedure Details} 
Participants began the study with onboarding, installing \system{}, optionally providing personal details, and completing a features tutorial. The subsequent \textbf{burn in} phase lasted 24 hours, during which \system{} collected data without showing suggestions, addressing the cold-start problem. The \textbf{active use} phase then spanned four days, with participants actively engaging with suggestions. During offboarding, we collected annotations and feedback on suggestions. We additionally conducted semi-structured interviews discussing participants' propositions, experiences, and privacy concerns.

\subsection{Analysis}
We first compute calibration and accuracy across all participants' labeled propositions, just like in our initial technical evaluation. In addition, the first author then conducted two rounds of qualitative open-coding across the transcripts. In the first round, the author labeled low-level themes across transcripts. In the second round, common themes were aggregated.

\subsection{Results}
We synthesize findings across both a quantitative analysis of \model{} accuracy and qualitative themes from our interviews. 

\paragraph{\model{}s remain accurate and calibrated in our end-to-end evaluation.} Mirroring results from our email evaluation, participants found sampled propositions to be accurate ($0.79 \pm 0.07$) and calibrated (Brier $ =0.28 \pm 0.04$), with error coming from underconfident predictions. Propositions that reflected factual statements about the user were often annotated or discussed with very little hesitation.
\begin{quote}
    \texttt{P1 is conducting research on health and behavior interventions.}
\end{quote}
\begin{quote}
    \texttt{P3 works on research related to civil and environmental engineering.}
\end{quote}
\begin{quote}
    \texttt{P4 is an employee at [ANON] company.}
\end{quote}
Participants also generally observed that these propositions were of higher confidence. For these propositions, participants expressed little surprise. 
\begin{quote}
    ``What's so special about this? It's just me.'' - P4
\end{quote}
In interviews, participants were quick to rationalize why these propositions were true, citing concrete websites, files, or messaging exchanges that explained how these propositions were generated. 

\paragraph{Maybe too accurate...} When reviewing propositions, participants were conflicted about generations that passed a judgement on values, skills, or norms. Consider the following propositions (all labeled correct by participants):
\begin{quote}
    \texttt{P1 is struggling with fixing bugs in their research project.}
\end{quote}
\begin{quote}
    \texttt{P2 may be experiencing stress or pressure due to a large number of unread emails and notifications.}
\end{quote}
\begin{quote}
    \texttt{P3 prioritizes communicating with their advisor over other people.}
\end{quote}
\begin{quote}
    \texttt{P4 is unhappy with their job.}
\end{quote}
Here, participants expressed a range of different opinions. P4 and P5 found these propositions both unserious and entertaining, ascribing little weight to them. P1, on the other hand, found them judgmental. In contrast, P2 immediately began self-reflecting:
\begin{quote}
    ``I felt so validated by that [proposition on stress]. I totally feel pressure when I get another email or notification. It's validating to see it notice this.'' - P2
\end{quote}
While P3 reflected on why they prioritized communication with their advisor,\footnote{The author of this paper also reflected on this.} they mentioned that only a subset of propositions were useful for reflection---propositions related to their abilities were not particularly constructive for them.
\begin{quote}
    ``I'd hide [propositions on ability], honestly. Even if they were true! Just show me the suggestions.'' - P3
\end{quote}
It was also trickier for participants to rationalize why their \model{} was making these inferences. For participants who enjoyed the self-reflection parts of \model{}, coming to their own conclusion was a boon; for others, this lack of transparency was annoying. We revisit self-reflection in the discussion (\S \ref{sec:disc_reflection}).

\paragraph{\system{} provides strong-to-excellent suggestions for all participants} All participants came to interviews especially excited about a suggestion generated by \system{}. On our 7 point Likert scale, 25\% of suggestions were ranked as strong (6) or excellent (7). P2 and P3 wanted to keep the system running on their computer after the study concluded (with some modifications to proposition visibility), while P2 and P5 took screenshots of a handful of suggestions to keep before offboarding. Suggestions that were rated highly often made extensive use of the underlying \model{}.
\begin{quote}
    ``It knew who my roommate was; what our budget was; where we were moving; that I was worried about this move. It worked backward from our move-in date, planned a schedule, and identified moving services. And half of my conversation with my roommate wasn't in English.'' - P1
\end{quote}
Overall, we found that useful suggestions generally fell into two main categories: immediate, low-level assistance, and opportunities where participants hadn't realized an AI model could be helpful.

\paragraph{In the moment help.} A subset of helpful suggestions were low-level, helping the user with what they were currently doing. P5, for example, was in the process of transferring email inboxes from their institution's email to Gmail, and never figured out how to enable notification on Gmail. \system{} generated a step-by-step guide---P5 was thrilled. In a similar flavor, P3 was designing a presentation for an upcoming quals exam. \system{} provided tips on slide transitions which P3 first tried and then adopted.

\paragraph{``I didn't even think about that!''} A more exciting class of suggestions generated by \system{} addresses needs that aren't explicitly signaled by the user. \system{} proactively generated a brainstorming outline for a framework P3 was considering using in their research, using tools P3 was comfortable with:
\begin{quote}
    ``What I typically like to do when I run into a new idea is create a barebones Overleaf document and outline [how the idea is] related to my work. I did not even tell this system anything, but it identified that I have this habit. And it created this entire outline---in LateX---of how I could write my paper in the context of this new framework I was checking out. I was like, wow.'' - P3
\end{quote}
P5 spent some time watching Twitch streams of games. For P5, \system{} identified specific characters of interest based on where and how long they paused the stream; and how often they replayed specific portions. Using this information, \system{} identified and surfaced new streamers that also played with the same character. P5 ended up exploring and watching content from \system{}'s recommended streamers.

\paragraph{Contextual agency.} 
Participants raised agency as a concern when interacting with \model{}, in terms of being able to control which contexts observations were turned into propositions or suggestions. P3 suggested that the system clarify with users when generating low-confidence propositions from observation, but acknowledged that the assumed propositions eventually did lead to useful suggestions.
\begin{quote}
    ``Maybe I wasn't familiar with PowerPoint, but if it asked and then used that to give me a suggestion I think I would've been happier with it.'' - P3
\end{quote}
For suggestions, \system{} did not replace what the user liked. P4, for example, really enjoys travel planning themselves, and their underlying \model{} included this fact. When generating suggestions, \system{} generated only ideas for travel instead of generating a full travel plan (something the system was capable of doing):
\begin{quote}
    ``I like that it gave me some ideas for adjusting my honeymoon schedule. It didn't try to redo the whole thing or make one from scratch.'' - P4
\end{quote}

\subsection{Errors and Boundaries}

\paragraph{Privacy, Privacy, and Privacy} Privacy guarantees were critical from the start. The two participants that were acquainted with the author explicitly mentioned this bias (which we revisit in the limitations). Recruitment alone required trust in the designer and that participants' \model{}s ran on self-hosted models.
\begin{quote}
    ``If I didn't know you and trust you, I would never install this thing.'' - P4
\end{quote}
Still, some participants (P1, P2) habituated and occasionally forgot that \system{} was even processing interaction data.
\begin{quote}
    ``It was intimidating at first. And I was also like shit like, do I need to be careful what I say? And then after a day I was like, whatever F it.'' - P2
\end{quote}
Others were keenly aware of the \system{}'s presence during the entire duration of the study (P3).
\begin{quote}
``It was just identifying things about me and understanding how I work, my work in general, and that was actually kind of scary.'' - P3
\end{quote}
Only after returning to the list of propositions would participants realize that \system{} was still active. On the other hand, P5 expressed no reservations during the entire study.

\paragraph{From General User Models to General Clippy Models?} All participants felt like \system{} often tried to recommend and execute on suggestions it was incapable of doing, a practical limit with agentic AI execution given today's models~\cite{bansal2024challenges, shao2024collaborative}. A good subset of suggestions (20.69\%) were labeled as on-track, but failed during execution. Still, some participants exacted joy from watching \system{} try and fail to execute.
\begin{quote}
``So my advisor’s been asking me to listen to this Claude Steele podcast---it’s been stuck at the top of my to-do list forever. I just never got around to it. Then this system gave me a nudge. It looked into the podcast and said it drew connections to my work. The outline was decent, but the connections were total garbage. Still, it got me to finally listen, and I ended up totally locked in, working on it for hours after. I almost prefer this, because it didn't take away any cognitive burden.'' - P2
\end{quote}
Still, this failure mode with \system{} reflects capability or tool-use limitations of the underlying LLM: \system{} did not explore the internet and ingest the podcast. Even simpler suggestions, such as scheduling reminders, also fail. \system{} claimed to have ``scheduled reminders'' for P5, allowing him to discreetly watch an NCAA game during his work hours. However, \system{} had no access to notification APIs, so execution failed entirely. 

\paragraph{Large \st{Language Models} Eager Beavers} Eagerness to help was a recurring theme amongst poorer suggestions. \system{} would jump the gun on generating suggestions while the user was in the process of finishing the suggestion itself. Or \system{} would try to cheat by suggesting something the participant had already completed. There's a fine line between suggestions that are in-the-moment helpful and suggestions that are too late.

\paragraph{Inferences on Sensitive Topics} P1 expressed some discomfort with \system{} making and saving inferences on sensitive topics. Here, \model{}s audit module incorrectly assumed that because P1 viewed and discussed topics with others, it was O.K. for it to save this information too. P1 mentioned that the integrity standards for \model{} should be strictly stronger than what their context implies---and that the Audit module should itself be auditable.

\paragraph{(Is) screen activity limited? Does it generalize?} 
All participants reflected on the generalizability of a \model{} trained on their computer use, and highlighted how events outside of their computer would not be captured by \system{}.  
\begin{quote}
    ``There were times that I was doing something on my phone and I was like, oh, I wish I could capture this too.'' - P2
\end{quote}
P3 also raised a similar point, explicitly reflecting on the privacy-capability trade-off of adding more context to a system like \system{}.
\begin{quote}
``There were some really great things that came out of there that I wouldn't have thought about. So how do I amplify that, but mitigate [seeing judgmental propositions]. I think I would need more than five days to use it. I would integrate some more personal work that I do.'' - P3
\end{quote}
We revisit this paradox in the discussion (\S \ref{sec:disc_privacy}).

\section{Discussion}
Across our technical evaluations and end-to-end deployment, we find that \model{} shows promise in various human-computer interaction scenarios. In our discussion, we focus on the implications of deploying \model{}s. We revisit how participants engaged and reflected with propositions, the \model{} privacy paradox, and the ethical and societal implications of widely deployed \model{}s.

\subsection{Reflecting on Propositions} 
\label{sec:disc_reflection}
We never intended or designed \system{} to serve as a self-reflection tool for participants. However, across both our email and end-to-end evaluation, many participants engaged deeply with the content of the propositions. For some participants, \model{} propositions were too candid: they explicitly constructed value judgements and made assessments about a user's priorities. This could be a side effect of our prompt---we explicitly prompt \model{} to make accurate and truthful assessments about the user. And these assessments are indeed useful for downstream suggestions generated by \system{}. Without an accurate user model, via an honest assessments of where the user might need help, downstream applications might fail to provide assistance. However, an accurate \model{} might enable a user to indulge in activities that, while they think are helpful, might not truly be helpful (e.g. cancelling advising meetings to make time for making/eating more ice cream).\footnote{This is a point of contention between the authors. Some of the authors are productive (the advisors), and would benefit greatly from an accurate user model. The first author thinks the ``Real Them'' would enjoy procrastinating.}

This dichotomy---between the ideal self and true self---is a well documented psychological phenomena~\cite{higgins1987self, rogers1995becoming}. When confronted with their \model{}, participants' reactions ranged from feeling judged to engaging in productive self-reflection.\footnote{More than once, participants likened \model{} to a BuzzFeed personality test.} Our work raises key design questions: should \system{} provide raw access to underlying user models, present sanitized versions, or hide propositions entirely? One approach might adapt the visible representation based on contextual factors, showing reflective content only when users are receptive to it~\cite{fleck2010reflecting}—similar to how our Audit module uses \model{} to make filtering decisions for privacy.

\subsection{The Privacy Paradox}
\label{sec:disc_privacy}
The privacy paradox observes that people often say they care very much about their online privacy, but willingly give it up in practice in exchange for a benefit~\cite{norberg2007privacy}. For most applications, this means giving up sensitive personal information such as location data or browsing habits to targeted advertisers via social media accounts. However, the \model{} is a much more general and, we argue, much more powerful user model. So, privacy behavior becomes a critical question. 

The more a \model{} sees, the more precise it becomes, and the stronger its downstream recommendations. Users can improve a \model{} by giving it anything that the model is capable of parsing. While several of the participants were initially focused on privacy risks, we observed that they  naturally came to the same conclusion and began voluntarily offering up additional data to the \model{}. In our end-to-end evaluation, two of our five participants wanted to continue using a version of \system{} following the study, brainstorming ways to give \model{} more data \emph{despite} initial privacy concerns.  

If one believes that sharing data with a private, local \model{} is safe and desirable, then disclosure becomes self-reinforcing: as users observe concrete improvements in the model's performance proportional to their data contributions, they may share more. However, if one considers the accidental disclosure risks or security risks of \model{}s, this loop could backfire. With this context, we outline a range of ethical and societal considerations in using \model{}s.

\subsection{Ethical and Societal Risks}

\paragraph{Persuasion, advertising, and surveillance} One risk we foresee lies in using \model{}s to target users, either through persuasion, advertisement, or surveillance. \emph{We strongly recommend that all \model{}s are trained and hosted on the end-users' computer or on personal infrastructure}. This is a guiding principle throughout our work---all models used to build a \model{} are open source, and infrastructure is managed by the research team. If external systems are ever granted access to the \model{}, access should be controlled and audited. We also foresee risks associated with third-parties attempting to train \model{}s, without consent, on a user's activity within their application. In these instances, obfuscation systems can generate interaction data to obscure genuine user behavior~\cite{nissenbaum2009trackmenot}.

\paragraph{Manipulating \model{}s} In our email evaluations, we found evidence of unintentional \model{} manipulation. Spam email, ingested by the \model{}, maliciously edited a \model{}'s underlying propositions. Instead of targeting users, attackers could craft messages that manipulate a user's \model{}. Any application that relies on a \model{} would therefore be affected. One possible mitigation for this risk would be for the \model{} to ask the user to confirm any inputs that seem out of character based on its understanding so far. Another would be to develop the equivalent of adblock or spam filtering techniques~\cite{sahami1998bayesian}) to proactively detect and intercept maliciously crafted content before it's processed by \model{}, or the model would need to include a component that reflects on whether someone is trying to trick it.

\paragraph{Bias} While participants did not raise concerns related to biased inferences, \model{}s build on biased models~\cite{dinan2020multi, weidinger2021ethical}. Analogously, recommendation systems have a long history of surfacing biased recommendations to end users~\cite{noble2018algorithms}. \model{}s effectively expand the surface area of recommendation systems. Carefully red-teaming and understanding the types of inferences constructed by \model{} across a larger set of adversarially constructed contexts is necessary before deployment.

\subsection{Limitations and Future Work}

\paragraph{Sampling bias} A challenge in recruiting participants for our end-to-end evaluation lies in trust. To this end, our recruitment occurred primarily at our institution and through word-of-mouth / snowball sampling.  Our institution’s IRB permitted up to 5 participants to mitigate the impact of any potential issues with \system{}. For now, we recommend considering our results as likely generalizing most strongly to technical users who have familiarity with AI.

\paragraph{Hallucinations and misattributions} Hallucinations remain an issue with current large models. While incorrect propositions are indeed appropriately calibrated, a handful are still confident and incorrect. \model{} would also occasionally hallucinate applications on a user's computer that didn't exist, or connect two unrelated propositions.

\paragraph{The screen is still a narrow proxy for context} While we see a wide range of the user via the GUM attached to a user's email or their screen, it’s not a complete picture---\model{}s are still far from building context through everyday action~\cite{dourish2004we}. If a user does something outside the scope of the \model{}, it will fail to generalize. However:

\paragraph{Better models will support longer contexts, be smaller, and extend beyond vision and language} \rev{Many components of the GUM pipeline work to mitigate the limitations of current-day LLMs, as simply putting everything in context is insufficent. Models forget parts of the context~\cite{liu2023lost} or yield generic propositions. Our retrieval step, for example, is required as models cannot reliably process extremely long context lengths. Explicit confidence verbalization is necessary as model logits are uncalibrated. However, we expect future work on modeling for GUMs (e.g. end-to-end learning parts of our pipeline) to subsume specific steps. We also expect future models to grow smaller in size and run entirely on-device, further mitigating privacy risks while reducing storage impact. Currently, the GUM saves about ~1GB of screenshots per day (just 5 MB of transcribed observations), which will accrue if deployed for longer. Another avenue for future work involves collecting minimal observational data for effective user models}. Finally, the current \model{} implementation relies on vision and language models, constraining the input space. Large models that ingest a wider range of modalities ~\cite{bommasani2021opportunities} along with diverse tool use could enhance \model{}'s capabilities. Already, advanced speech models can discern a user's tone, while healthcare models interpret various sensor signals---all of which in principle could integrate into the \model{}.

\section{Conclusion}
We introduce GUMs, computational models that learn rich representations of user context by observing everyday behavior. GUMs transform unstructured interactions into confidence-weighted propositions, continuously refining their inferences as new evidence accumulates. \model{}s open up a range of possibilities, from grounding language models in observed user behavior to enabling proactive systems that can act on a user's behalf. We demonstrate the utility of GUMs through \system{}, an interactive assistant that surfaces helpful suggestions drawn directly from personal context and attempts to complete them. In evaluations, GUMs produce rapid, calibrated, and accurate inferences about users. We argue that GUMs provide a framework for moving beyond siloed application-specific user models toward a \emph{general} user model---one that understands who you are across the many contexts of your life.

\begin{acks}
We thank Michael Y. Li, Jensen Gao, Suvir Mirchandani, Lindsay Popowski, Ryan Louie, Beleicia Bullock, Will Held, Dora Zhao, Tiziano Piccardi, Michelle Lam, and Carolyn Zou for helpful discussions and feedback. We also thank members of the Stanford HCI and NLP groups for feedback in general and for testing both \model{} and \system{}. We don't thank Omar's ice cream making hobby for filling this paper with examples that might make the reader hungry. Omar Shaikh is supported by the HAI-HPI program and Joon Park is supported by the Microsoft Research fellowship. We also thank SCBx, AXA, TRI, American Express, Banco Itaú, Ford, Hanwha, and Google, as well as ONR grant N000142412532 for helping fund our work.
\end{acks}

\bibliographystyle{ACM-Reference-Format}
\bibliography{base}

\clearpage
\appendix
\section{Code}
A demo of \system{} and our open source package for \model{} are available at \url{https://generalusermodels.github.io}

\section{Abridged Prompts}

Here, we outline prompts from our work in more detail. We discuss prompts used in the underlying \model{} and the instantiated system, \system{}.

\subsection{\model{} prompts}

\subsubsection{Calibration}
\label{appdx:calibration}
\model{} is responsible for generating confidence scores alongside each proposition. One method to do this is to directly look at the logprobs on the completion from the LLM. However, this estimate for instruction-tuned LLMs is often miscalibrated: LLMs are more confident than they really are. An alternative approach involves prompting LLMs for a verbalized confidence score~\cite{tian2023just}. This entire process occurs during proposition generation, after we've already generated the reasoning and the proposition:
\begin{quote}
\texttt{observation: [screenshots of the user switching between Overleaf and YouTube]}
\end{quote}
\begin{quote}
\texttt{proposition\_reasoning: \textbf{``The user appears distracted, switching focus between an ice cream recipe video and typing intermittently in an Overleaf window.''}}  
\end{quote}
\begin{quote}
    \texttt{proposition: \textbf{``User periodically views ice cream recipes while writing.''}}   
\end{quote}
Conditioned on the above, we elicit a support score (1-10), which serves as our confidence measure. An abridged version of our prompt is below:
\begin{quote}
    \texttt{Generate a support score that captures how much evidence you have to support the generated propositons. Be conservative in your support estimates. Just because an application appears on the screen does not mean they have deeply engaged with it. They may have only glanced at it for a second, making it difficult to draw strong conclusions.} 
\end{quote}
\begin{quote}
    \texttt{Assign high support scores (e.g., 8-10) only when the transcriptions provide explicit, direct evidence that the user is actively engaging with the content in a meaningful way.}   
\end{quote}
Using the above, we generate a support score:
\begin{quote}
    \texttt{support: \textbf{10}}
\end{quote}
We elicit conservative confidence scores, as instructed in our prompt. \model{} is overall well-calibrated and does indeed generate conservative support scores. We suspect that \model{}'s calibration could be improved either by finetuning or through better prompting---we leave this for future work.

\subsubsection{Reranker}
\label{appdx:reranker}
To marshall the \model{}, we need to be able to retrieve relevant propositions. A challenge here requires \emph{re-ranking} outputs---many of the retrieved propositions may be irrelevant. To this end, we classify retrieved results using the following (abridged) prompt:

\begin{quote}
\texttt{Classify the similarity between two propositions as:}
\\
\\
\texttt{(A) HIGHLY RELATED - practically or exactly the same.}
\\
\texttt{(B) SOMEWHAT RELATED - similar idea or topic.}
\\
\texttt{(C) DIFFERENT - fundamentally unrelated.}
\\
\\
\texttt{Proposition A:
\{proposition\_a\}}

\texttt{Proposition B:
\{proposition\_b\}}
\\
\\
\texttt{Respond ONLY with: A, B, or C.}
\end{quote}
Based on the classification outputs, we skip, revise, or add to the \model{}.
\subsection{\system{} prompts}

\subsubsection{Mixed Initiative Interaction}
\label{appdx:mii_prompt}

\system{} generates suggestions to show users, but we can't \emph{show} users all suggestions. To this end, we need to filter what suggestions worth showing. We instantiate mixed-initiative interaction, but this approach requires estimating both the probability of a suggestion being useful and its utility to the user. Here, we use the \model{} to generate both the probability of the user selecting the suggestion, and the costs and benefits for the user. We use the following (abridged) prompt:

\begin{quote}
\texttt{Evaluate each suggestion (1--10 scale):}\\[6pt]
\texttt{1. \textbf{Benefit}: How helpful is assistance for \{user\_name\}?}\\
\texttt{(1 = not beneficial, 10 = highly beneficial; consider simplicity, genericness, user's current actions, urgency.)}\\[6pt]
\texttt{2. \textbf{False Positive Cost}: How disruptive would unsolicited assistance be?}\\
\texttt{(1 = not disruptive, 10 = highly disruptive; consider user's workflow and focus.)}\\[6pt]
\texttt{3. \textbf{False Negative Cost}: How critical is assistance if genuinely needed?}\\
\texttt{(1 = no impact, 10 = significant negative impact; consider potential setbacks without help.)}\\[6pt]
\texttt{4. \textbf{Decay}: How quickly does the suggestion's benefit diminish over time?}\\
\texttt{(1 = immediately obsolete, 10 = remains useful long-term; consider urgency and task deadlines.)}
\end{quote}
As context, we provide the suggestion, propositions, and underlying observations in the prompt.

\subsubsection{Tool Use}
\label{appdx:tool_needed}
Finally, \system{} relies on external tools for execution. Here, we use a prompt that selects a subset of tools worth using (with the suggestion / observation in context). An abridged version is below:

\begin{quote}
\texttt{\textbf{What tools should you use?}}

\texttt{Here are the tools you have at your disposal:}\\[1pt]

\texttt{\textbf{llm} (e.g. no tools)}\\
\texttt{- Generate responses directly without using any tools.}\\[1pt]

\texttt{\textbf{search}}\\
\texttt{- Search for topics online and provide citations.}\\
\texttt{- Use liberally for latest internet information.}\\[1pt]

\texttt{\textbf{filesystem}(parameter: (str) filename)}\\
\texttt{- ONLY use if the file certainly exists on the user's computer.}\\
\texttt{- Helpful when viewing the entire file aids user assistance.}\\[1pt]

\texttt{\textbf{reasoning}}\\
\texttt{- For challenging coding/math problems requiring deeper thought.}\\[1pt]

\texttt{\textbf{image}(parameter: prompt)}\\
\texttt{- Generate an image given a specific prompt.}\\[1pt]

\texttt{Generate a list of useful tools with parameters in JSON format:}
\end{quote}
\section{Survey Questions}
\label{appdx:survey}
Below, we outline all the survey questions asked to participants in the email study.
\begin{enumerate}
    \item Demographic Information: Name, Age, Gender, Race/Ethnicity, Profession, Email \hfill (Short answer)
    \item How often do you typically check your email? \hfill (Multiple choice: frequency options)
    \item How many different email accounts do you use regularly? \hfill (Multiple choice: 1, 2, 3, 4+)
    \item What do you primarily use this specific email for? \hfill (Checkboxes: use cases)
    \item Briefly describe the type of content that shows up in this email inbox. \hfill (Paragraph)
    \item How accurately do you feel the propositions shown to you were in reflecting what's in your email? Why do you think they are or aren't accurate? \hfill (Paragraph)
    \item Overall, how accurate were the propositions with high confidence with respect to the context in the email? \hfill (7-point Likert)
    \item Overall, how relevant were the propositions to the context in the email? \hfill (7-point Likert)
    \item Any thoughts about specific propositions? If so, which ones and why? \hfill (Paragraph)
    \item Which data sharing generation was most accurate? \hfill (Multiple choice: First, Second, Third)
    \item How much do you agree with the answers to the best data sharing response you picked earlier? Rate these based on accuracy. \hfill (7-point Likert)
    \item Did you feel the propositions themselves (the text) respected your context and privacy? Why or why not? \hfill (Paragraph)
\end{enumerate}

\section{Interview Questions}
\label{appdx:interview}
For our end-to-end evaluation, we conduct an hour-long semi-structured interview. We guide the interview using the questions below. Since the interview is unstructured, we also ask followups.
\subsection{Overall Experience}
\begin{enumerate}
    \item What were your initial expectations for using Gumbo?
    \item How would you describe your overall experience using this?
    \item Was there anything that fell short of your expectations?
\end{enumerate}

\subsection{Propositions}
\begin{enumerate}
    \item How well do you think the propositions understood you?
    \item Can you describe how you felt during looking at the propositions?
    \item Was there anything about the propositions that you particularly liked?
    \item Did you edit any of the propositions?
    \item How accurate did you find the propositions to be?
    \item Were there any propositions that you strongly disagreed with? What was wrong?
    \item Did the propositions reveal something to you that you weren’t aware of?
\end{enumerate}

\subsection{Suggestions}
\begin{enumerate}
    \item How well do you think the suggestions understood your goals and concerns based on your interactions?
    \item Did the system offer any insights or recommendations that you found useful?
    \item Did the system offer any insights or recommendations that you did not find useful?
    \item Did you act on its advice?
\end{enumerate}

\subsection{General Reflections}
\begin{enumerate}
    \item If you were able to continue using this system over time, what would you want to use it for?
    \item What features would you add to improve this?
    \item Would you recommend this application to others? What would you tell them about it?
    \item Is there anything else you would like to share about your experience that we haven't already asked?
\end{enumerate}

\section{Screen Observer Outputs}
\label{appdx:screen_outputs}

Our screen observer provides transcriptions that the \model{} uses to create propositions. The raw update is entirely in text. The first part of the update is a text-based transcription of the screen (abridged output below), where the first author was editing the Swift frontend for \system{}:
\begin{quote}
\texttt{Running Application:} \\
\texttt{- Xcode (window title: Running Horizon)} \\

\texttt{Open Tabs in Xcode:} \\
\texttt{- Horizon > main} \\
\texttt{- OnboardingManager} \\
\texttt{- FrameProcessingService} \\
\texttt{- AddPropositionPopup} \\
\texttt{- UserModelPage} \\
\texttt{- DataModelManagerAPI} \\
\texttt{- AppDelegate} \\

\texttt{Open File in Editor:} \\
\texttt{- File: DataModelManager.swift} \\
\texttt{- Function: fetchNewProjects(completion:)} \\
\texttt{[omitted code here]}
\end{quote}
The second part of the transcription is a description of actions the user takes
across the past 10 unique frames (anonymized example below):
\begin{quote}
\texttt{In Figma, the user is designing a UI mockup titled "User Model" that displays structured propositions about a user along with confidence values. They are refining grouped elements and layout, including recommendation cards like “Suit Rentals in Chicago” and “Flights to Chicago”.}
\\
\\
\texttt{In Overleaf, the user is editing a LaTeX document for a paper titled “General User Models.” They are working in draft.tex, using input commands to include section files, and referencing the Figma UI figure in “Figure 1.”  A warning is possibly due to a formatting issue.}
\\
\\
\texttt{In Xcode, the user is developing or debugging a Swift-based application named “Horizon.” They are working in the file DataModelManager on a function fetchNewProjects(completion:) that appends suggestions and sends notifs if certain utility thresholds are met. The debug console shows a  backend communication error (404 no new updates from the backend) and a missing  symbol warning.}
\end{quote}

\end{document}